\documentclass[nofootinbib,onecolumn,preprintnumbers,amsmath,amssymb]{revtex4}
\usepackage{graphicx}% Include figure files
\usepackage[latin1]{inputenc}
\usepackage{dcolumn}% Align table columns on decimal point
\usepackage{bm}% bold math
\usepackage{epsfig}
\usepackage{tabularx}
\usepackage{color}
\usepackage{tikz}
\usepackage{nameref,hyperref}
\usepackage{comment}
\usepackage{amsmath,amsthm}
\def\checkmark{\tikz\fill[scale=0.4](0,.35) -- (.25,0) -- (1,.7) -- (.25,.15) -- cycle;}

\begin{document}

%\preprint{\texttt{The order of the authors is currently random -- don't pay attention to it}}

\title{A combinatorial framework to quantify peak/pit asymmetries in complex dynamics}% Force line breaks with \\

\author{Uri Hasson$^{1,2}$, Jacopo Iacovacci$^{3,4}$, Ben Davis$^{1}$, Ryan Flanagan$^5$, Enzo Tagliazucchi$^6$, Helmut Laufs$^{7,8}$ and Lucas Lacasa$^5$}
%\email{l.lacasa@qmul.ac.uk}
\affiliation{$^{1}$Center for Mind and Brain Sciences, University of Trento (Italy)\\
$^{2}$Center for Practical Wisdom, The University of Chicago, Chicago (USA)\\
$^{3}$Department of Surgery and Cancer, Division of Computational and Systems Medicine, Imperial College London, London SW7 2AZ (United Kingdom)\\
$^{4}$The Molecular Biology of Metabolism Laboratory, The Francis Crick Institute, London NW1 1AT (United Kingdom)\\
$^{5}$School of Mathematical Sciences, Queen Mary University of London, E14NS London (United Kingdom)\\
$^{6}$Netherlands Institute for Neuroscience, Meibergdreef 47, 1105 BA Amsterdam-Zuidoost (The Netherlands) \\ 
$^{7}$Department of Neurology and Brain Imaging Center, Goethe University Frankfurt am Main (Germany). \\
$^{8}$Department of Neurology, University Hospital Kiel, Kiel (Germany)}%

\email{l.lacasa@qmul.ac.uk | uri.hasson@unitn.it}

\date{\today}% It is always \today, today,

\begin{abstract}
We explore a combinatorial framework which efficiently quantifies the asymmetries between minima and maxima in local fluctuations of time series. We firstly showcase its performance by applying it to a battery of synthetic cases. We find rigorous results on some canonical dynamical models (stochastic processes with and without correlations, chaotic processes) complemented by extensive numerical simulations for a range of processes which indicate that the methodology correctly distinguishes different complex dynamics and outperforms state of the art metrics in several cases. Subsequently, we apply this methodology to real-world problems emerging across several disciplines including cases in neurobiology, finance and climate science. We conclude that differences between the statistics of local maxima and local minima in time series are highly informative of the complex underlying dynamics and a graph-theoretic extraction procedure allows to use these features for statistical learning purposed. 
\end{abstract}

 \maketitle

\section{\textsf{Introduction}}
A major challenge in studying temporally unfolding natural systems is making sense of data that are noisy, reflect processes at local and global scales, and are likely non-stationary. This is exemplified in diverse domains in the natural sciences where theoretically-driven research aims to describe how system organization and interaction dynamics relate to produce time series features \cite{NTSA,ecoTSA}. Scientists therefore require time-series descriptors that are: highly informative while still offering a good compression, relatively robust against instrument induced fluctuations, and (particularly in neurobiological domains) invariant to individual differences that exist between samples.\\
\noindent Particularly important for research across disciplines including neurobiology, finance and earth sciences is that such descriptors differentiate the contribution of oscillatory processes from that of point-like impulses that are inherently non-oscillatory.  This is patently apparent when considering spectral power measures, where both oscillatory processes and regularly spaced point-like events produce similar signatures.  Additionally, the structure of point-like events could itself reflect different processes that determine local minima and maxima in a time series. 
% moved this sentence to Neurobio below where it is developed and better integrated. For instance, studies of brain function have identified systems where the magnitude of activity peaks tracks sensory features (e.g., pitch), but activity pits are impacted by the processing of other sorts of stimuli, or modulated by overall attention (e.g. activity in auditory cortex). 
This translates into a need to quantitatively measure the anisotropy of a fluctuating signal in a manner that captures to the difference in the structure of peaks (local maxima) and pits (local minima). 
Departing from analytic approaches based on consideration of an entire time series -- captured by variants of spectral analysis, entropy, or non-linear signatures -- one that contrasts 
structure within local peaks and pits is certainly needed, and we present such an approach here.\\
%as we develop here produces a fundamentally different
%understanding.  Among other properties, it differentiates oscillatory, random and chaotic
%series and produces summary statistics that describe the temporal scales for which peak
%and pit progressions differ.}\\

%We shall develop this point in detail, because progress in neuroscience demonstrates how scientists can be lead down the garden path in its absence.\\
To further motivate the aforementioned theoretical challenge, we shall discuss the issue in some concrete disciplines. In neurobiology, and more specifically within research examining the brain's spontaneous mode of operation during wakeful rest (``resting state processes"), it has long been known that
Blood-oxygen-level-dependent (BOLD) time series in the human brain show strong power at low frequencies ($<0.1$Hz) and that such frequencies underlie resting-state
connectivity \cite{Cordes01}. This has produced an entrenched view of natural brain dynamics: these dynamics are taken to reflect slow, synchronous oscillations at low frequencies. However, more recent research is shedding light on the problems with such descriptions. For instance, \cite{Petridou13} identified strong and highly infrequent spontaneously occurring spike-like events within recordings of resting state brain activity. They showed that removing these events from the time series reduced low-frequency power by as much as $60\%$ in certain brain areas. This removal also reduced correlations within brain networks by as much as $50\%$. Other work suggests that such infrequent spontaneous events carry much weight in explaining important phenomena related to brain function. In particular, Taglizaucchi et al.  \cite{Tagliazucchi16} showed that well-studied connectivity networks in the human brain can be accurately reconstructed even keeping \emph{only} $1\%$ of the data in each time series (for related work, see \cite{Liu13}). \\
% These examples show that analyses that suggest that brain activity is largely oscillatory at low frequencies are also consistent with activity profiles that are driven by infrequent discrete events. Trains of short-lived and temporally spaced impulses can masquerade as low-frequency fluctuations, while actually reflecting a completely different activity regime. UH: I'm deleting this to offer a better transition to the main idea
In parallel with the shift towards consideration of the role of rare, relatively extreme events, other work began examining in more detail the specific features of both local maxima and local minima in neurobiological time series. {Many neurobiological time series, including those measured by fMRI scanners, do \textit{not} have a natural minima where the measured signal is zero.} From the neurobiological perspective, the need to understand minima and maxima in such cases emerges naturally within any model in which oscillatory patterns or power spectra are not the only source of information on brain activity.\\ 
To illustrate, in the auditory system, activity peaks are known to track physical features of the input (e.g., frequency or amplitude), so that external stimuli produce well defined steady-state responses (peaks in activity spectra) that track time-varying features of auditory inputs (e.g., \cite{Patel00}). However, within the same auditory system, activity pits may be further impacted by other factors, including "dampening" indcued by visual processing, the level of  overall attention paid to the auditory stimuli, or the degree of engagement in memory maintenance. This is a simple example of how local maxima and local minima may provide information about different processes. Additionally, studies of resting state neuroelectric responses in the human brain have distinguished between modulations of oscillatory peaks from modulation of oscillatory troughs. That line of work has documented a difference between the variance of peaks and pits in resting-state time, (Amplitude Fluctuation Asymmetry \cite{Mazaheri10}), linking these to activity in visual areas, and suggesting this asymmetry derives from unbalanced forward vs. backward propagating currents within dendrites. Studies of functional Magnetic Resonance Imaging (fMRI) have also examined asymmetries in resting-state activity. These studies first identified all local minima and maxima in each time series and then contrasted the respective variances of the sets of local maxima and local minima.  This variance asymmetry within spontaneous brain activity distinguishes adults from children \cite{Davis14} and differentiates between wakefulness and sleep stages \cite{Davis16}. As noted by Mazehari and Jensen \cite{Mazaheri10}, ``the amplitude fluctuations of oscillatory activity are conventionally viewed as being symmetric around zero'', but as we summarized above, emerging findings show that these conventions require revision, and importantly -- that new and precise analytic tools are needed to quantify features of asymmetry within local minima and maxima. As we later show, amplitude asymmetry might be less than ideal in identifying such dynamics.\\

\noindent The need to capture and quantify possible asymmetries between the local maxima and minima of time series is not just inherent in neurobiological processes. Let us consider the time evolution of some financial index $x(t)$ (e.g. S\&P500) which represents the aggregate, collective evolution of the interaction of a number of financial assets over time. In quantitative finance, it is a well-known empirical fact that qualitatively different dynamics operate microscopically when $x(t)$ is on average increasing with respect to when this index is in a sustained decrease (which is equivalent to say that $-x(t)$ grows). In the former case,  market is usually stable, pairwise correlation between the constituting assets is generally low (the system is said to be close to equilibrium \cite{Flanagan2016}) and risk perception is low, leading to a time series with low volatility. Conversely, a situation where $x(t)$ decreases is indicative of a market under stress, where the correlation of the constituent assets grows due to common factors. As a result, any small and uncontrollable fluctuation can easily propagate throughout the system, hence the dynamics display larger volatility (larger uncertainty). The role that these two  different market dynamics is playing can be therefore examined by analysing series statistics under index inversion $x(t)\to -x(t)$, something that we will show is tightly related to asymmetries between peak and pit statistics.
\\

\noindent Keeping in mind the aforementioned desiderata and as well as findings highlighting the importance of
discrete events, our aim here was to develop a general method for the efficient quantification of peak/pit asymmetry targeted at understanding the role
of local, non-oscillatory processes. After a thorough validation of such a methodology, our aim was then to apply it in a wide range of settings, including neurobiological, financial, climate time series and beyond. We capitalized on a recent general approach to the description of time series that provides information about both local and global temporal features, without assuming neither stationarity nor oscillations at any temporal scale. 
This approach originates on the notion of a visibility graph \cite{Lacasa2008, Lacasa2009,multivariate}, a mapping that converts an ordered sequence of $N$ real-valued data (e.g. time series) into a graph of $N$ nodes, where each time point is mapped into a node and two nodes are connected if certain geometric and ordering criteria hold amongst the data (see figure \ref{fig:VGHVG} and Methods).  Visibility Graphs were introduced with the aims of using the tools of Network Science \cite{Newmanbook} to describe the structure of time series and their underlying dynamics from a combinatorial perspective (other proposals for graph-theoretical time series analysis can be found in \cite{Kurths2017} and references therein). Research on this methodology has been primarily mathematical (elaborating on mathematical methods \cite{severini, Luque2016,nonlinearity,motifs} to extract rigorous results on the properties of these graphs when associated to canonical models of complex dynamics \cite{epl, jns, pre2013,quasi}. In practice, this method can be used as a feature extraction procedure for constructing feature vectors for statistical learning purposes (see \cite{Shao2010,Ahmadlou2010,Ahmadlou2012,Bhaduri2016,Sanino,meditation_VG} for just a few examples in the life sciences or \cite{physics3,fluiddyn0,fluiddyn1,fluiddyn2,physics2,suyal,Zou,physio1} for other applications in the physical sciences).\\
%This philosophy indeed provides a computationally effective way for studying complex neurobiological data because it is sensitive to temporal patterns at very local scales (on the scale of 2-4 measured time points). 
%This is because
%the node degree of each vertex (time point) \emph{just} reflects the vertice's relative magnitude in a local
%neighborhood.\\ 
Importantly, a fundamental property is that the conversion from time series to
visibility graphs is invariant under several transformations that map onto to common
nuisance (e.g., instrument) effects which are typical in neurobiological time series and beyond. These include linear trends,
amplitude modulation on longer scales, or variations (slowing-down, speeding-up) in the rate of the process in question. Consequently, these invariances result in  more efficient combination of information across
measurements after transforming a time series into a visibility graph. Visibility graphs provide informative features both at the local level of single
vertices and at the global level, where distributions of vertex features are described.
Such global features reflect, for instance, the self-similarity of fractal time series \cite{epl}, estimation of entropy production due to time irreversibility \cite{epjb}, discrimination between noise and chaos \cite{toral,ravetti}, etc. Interestingly, previous works report that visibility graph features capture both linear and nonlinear information of the dynamical process and thus extend above and beyond standard power spectrum-like measures which only capture linear correlations.\\
%This method has been applied to EEG data in classifying clinical state
%(doi:10.1016/j.physa.2012.04.025) or wakefulness (10.1109/JBHI.2014.2303991), as %well as
%to classification of cardiac series taken from different populations [REF]. 
Additionally, because visibility graphs can provide insights into local, non-oscillatory processes \cite{motifs}, they go
beyond the information provided by `global' measures that summarize time
series in a single parameter \cite{NTSA} (series entropy, fractal dimensions, power spectrum
or even methods that are sensitive to similar (repetitive) motifs on multiple time scales (e.g., multiscale entropy methods)).\\

\noindent Equipped with the notion of visibility graph as a starting point, we explore here a systematic extension to that method, which is designed to satisfy the desiderata outlined above and capture different dynamics that may determine peaks and pits in a given signal. As indicated above, we are interested in a measure of asymmetry that can be applied to time series that do not have a natural minimum, which can be combined across measurements, and which is relatively robust to noise. \\
The rest of the paper proceeds as follows: in the Methods section we introduce the theoretical formalism, along with a theoretic analysis and validation for synthethic processes. For simple (stochastic and deterministic) processes we show that this methodology can capture peak/pit asymmetries with similar performance to that state of the art indicators, however for more complex processes involving combination of different dynamics and scales we show that this methodology outperforms state of the art approaches. We confirm such findings with extensive numerical simulations and rigorous results on concrete, canonical complex dynamics.
In the results section, we first show that this method offers novel descriptions for spontaneous brain activity in humans and differentiates between states of consciousness. For financial time series, it captures important features of financial regimes linked to major events in stock markets. We finally explore the application of this framework for extensive climatic data.
  
\section{\textsf{Methods}}
\subsection{\textsf{Graph-theoretical framework for peak/pit asymmetry quantification}}
\noindent {\bf Peak and Pit subsets. }
Let us consider a real-valued time series of size $N$, ${\cal X}=\{x(t)\}_{t=1}^N$. The traditional step to get access to peak and pit statistics is to define two ordered sets, namely $\it{peak}=\{x(t)| x(t+1)<x(t), x(t-1)<x(t)\}$ and $\it{pit}=\{x(t)| x(t+1)>x(t), x(t-1)>x(t)\}$. 
The hypothesis is that the statistics of these sets, and its difference, carry relevant information on local fluctuations of $\cal X$ and can be used as a feature for making diagnostics. Mathematically, differences in the statistics of {\it peak} and {\it pit} can be related in principle to two scenarios, namely: (${\cal S}_1$) different marginal distributions, and (${\cal S}_2$) different correlations (i.e. {\it peak} and {\it pit} might have similar marginals but different temporal correlations). Additionally, (${\cal S}_3$) characteristic cross-correlations between {\it peak} and {\it pit} can also be informative (e.g. {\it peak} and {\it pit} might have similar marginal distributions but say, fluctuate in an anti-correlated way with one another). Among the plethora of possible descriptors, one should be able to identify and separate what scenarios (${\cal S}_{1-3}$) the measures are addressing. As a matter of fact, asymmetries between peak and pit statistics have only been addressed relatively recently, and the state of the art statistical tests e.g. Amplitude Fluctuation Asymmetry (AFA \cite{Mazaheri10}) or Amplitude Variance Asymmetry (AVA \cite{Davis14}) essentially considers scenario (${\cal S}_1$) by comparing the variances of these marginals (through the logarithm of the so-called  Variance Ratio $\text{VR}= \sigma^2(\text{peak})(t)/\sigma^2(\text{pit}(t))$ in the case of AVA, where $\sigma^2(X)$ is the variance of the random variable $X$). 
As these measures only considers the one-point marginals of each set, they cannot give us particular insights on scenarios (${\cal S}_2$) or (${\cal S}_3$), and therefore disregard these aspects. However it is a computationally simple statistic, and probably because of its simplicity the quantity $|\log(\text{VR})|$ is currently used to assess the similarity between peak and pit statistics as mentioned above when describing prior applications to neurobiological time series. In addition, the degree to which AVA or AFA in any given time series is driven by extreme points \cite{Amor} is unclear and necessitates manual examination. In practice, it appears there may be a relation between the AVA of brain time series and the skewness of the empirical distribution \cite{Amor}, though analytically AVA and skewness are independent as the same distribution can produce time series with markedly different AVA values depending on specific sampling.
Incidentally, note that the quantity $|\log(\text{VR})|$ fulfils the axioms of a metric only in the case where $\log(\text{VR})$ is positive\footnote{This can be seen from the triangle inequality: take $X$, $Y$, $Z$. For $\sigma^2(X)>\sigma^2(Y)>\sigma^2(Z)$ we can drop the absolute values and the triangle inequality saturates 
$$\log\bigg(\frac{\sigma^2(X)}{\sigma^2(Y)}\bigg) + \log\bigg(\frac{\sigma^2(Y)}{\sigma^2(Z)}\bigg)= \log\bigg(\frac{\sigma^2(X)}{\sigma^2(Z)}\bigg)$$
Take however $\sigma^2(X)>\sigma^2(Y)$, $\sigma^2(Y)<\sigma^2(Z)$. In that case the triange inequality is not satisfied in general, as we have
$$\log\bigg(\frac{\sigma^2(X)}{\sigma^2(Y)}\bigg) - \log\bigg(\frac{\sigma^2(Y)}{\sigma^2(Z)}\bigg)= \log\bigg(\frac{\sigma^2(X)\sigma^2(Z)}{\sigma^2(Y)\sigma^2(Y)}\bigg)$$ which in general is not larger or equal than $\log\bigg(\frac{\sigma^2(X)}{\sigma^2(Z)}\bigg)$.}, so other proposals that not only are able to quantify (${\cal S}_{1-3}$) but which might rely on more solid mathematical grounds than AVA are needed. An obvious strategy could directly define similarity measures not only from single-point distributions of the {\it peak} and {\it pit} sets (mean, variance, etc), but explore higher-order joint distributions of these datasets (two-point, three-point and in general m-point distributions of strings of size $m$). This however is not likely to be efficient in practice, as high-order statistics usually require access to very long time series (usually the length of the observed time series is required to grow exponentially with $m$).\\ 
Furthermore, note that by construction $(\it{peak} \cup \it{pit})\subset {\cal S}$, meaning that this decomposition is lossy: in a generic fluctuating signal there are data which are neither in the peak nor in the pit set, so a priori important features of the local fluctuations might be lost if one only looks at the peak and pit sets and discard intermediate data. 
An alternative, which we explore in what follows, is to be able to extract high-order features from peak and pit local neighborhoods by projecting the full signals into an appropriate topological space.\\

\begin{figure}
\centering
\includegraphics[width= 13 cm]{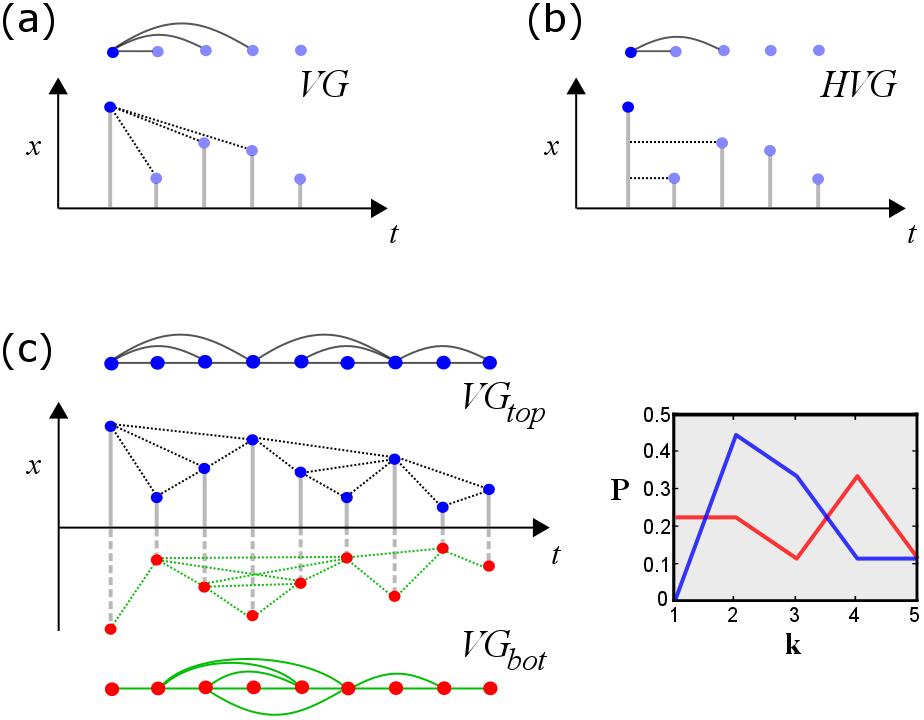}
\caption{(Panels a and b) Sample time series and associated natural visibility (VG, panel a) and horizontal visibility (HVG, panel b) linking criteria. Each mapping is invariant under a certain number of transformations (VG is invariant under affine transformations $x\to ax+b$, $a>0$, HVG is invariant under monotonic transformations $x\to f(x), \ f(\cdot) $ an order-preserving function). (Panel c) Sample time series and extraction of the top natural visibility graph (black) and bottom natural visibility graphs, which characterise the visibility structure of local maxima (top) and local minima (bottom) respectively. Note that $G_{\text{top}}$ coincides with a standard visibility graph of series $x(t)$, and $G_{\text{bot}}\{x(t)\}\equiv G_{\text{top}}\{-x(t)\}$, that is, characterization of local minima is achieved by extracting the visibility graph from the inversed series $-x(t)$. One can extract features from both $G_{\text{top}}$ and $G_{\text{bot}}$ (here, a cartoon of the degree distribution $P(k)$) and compare these as a proxy to the comparison between local minima and local maxima statistics.}
\label{fig:VGHVG}
\end{figure}

\noindent {\bf Visibility graphs and top-bottom VG/HVG asymmetry ($\Delta$VGA). }
Consider again a time series ${\cal X}=\{x(t)\}_{t=1}^N$. The so-called natural visibility graph (VG) \cite{Lacasa2008} is extracted from the series by associating a different node to each datum, and linking every two nodes $i$ and $j$ with an edge if the following convexity criterion holds between the associated time series data: 
\begin{center}
$x(k) < x(i)+[x(j)-x(i)]\displaystyle\frac{k-i}{j-i}, \ \forall k: i<k<j.$
\end{center}
Related to this graph, the so-called horizontal visibility graph (HVG) \cite{Lacasa2009} is a subgraph of VG, obtained by applying a similar procedure with a slightly different linking criterion which only relies on the ordering of the data: 
\begin{center}
$x(k) < \inf\{x(i),x(j)\}, \ \forall k: i<k<j.$
\end{center}
The theory of VG/HVG has been intensively used in recent years, not only to display a combinatorial representation of complex dynamics but also as a computationally efficient way of extracting informative topological features from empirical time series for statistical learning purposes. Among other properties, features of these graphs include a simple way to estimate the Hurst exponent in fractional Brownian motion \cite{epl} or the ability to easily distinguish fully chaotic processes from uncorrelated stochastic ones \cite{toral,ravetti,motifs} (both having identical flat power spectra). The Lyapunov exponent of a chaotic map can also be quantitatively obtained from the graph's block entropies \cite{wolfram}. All in all, visibility graphs extracts valuable information from a time series on both linear and nonlinear level.\\

\begin{figure}
\centering
\includegraphics[width= 9 cm]{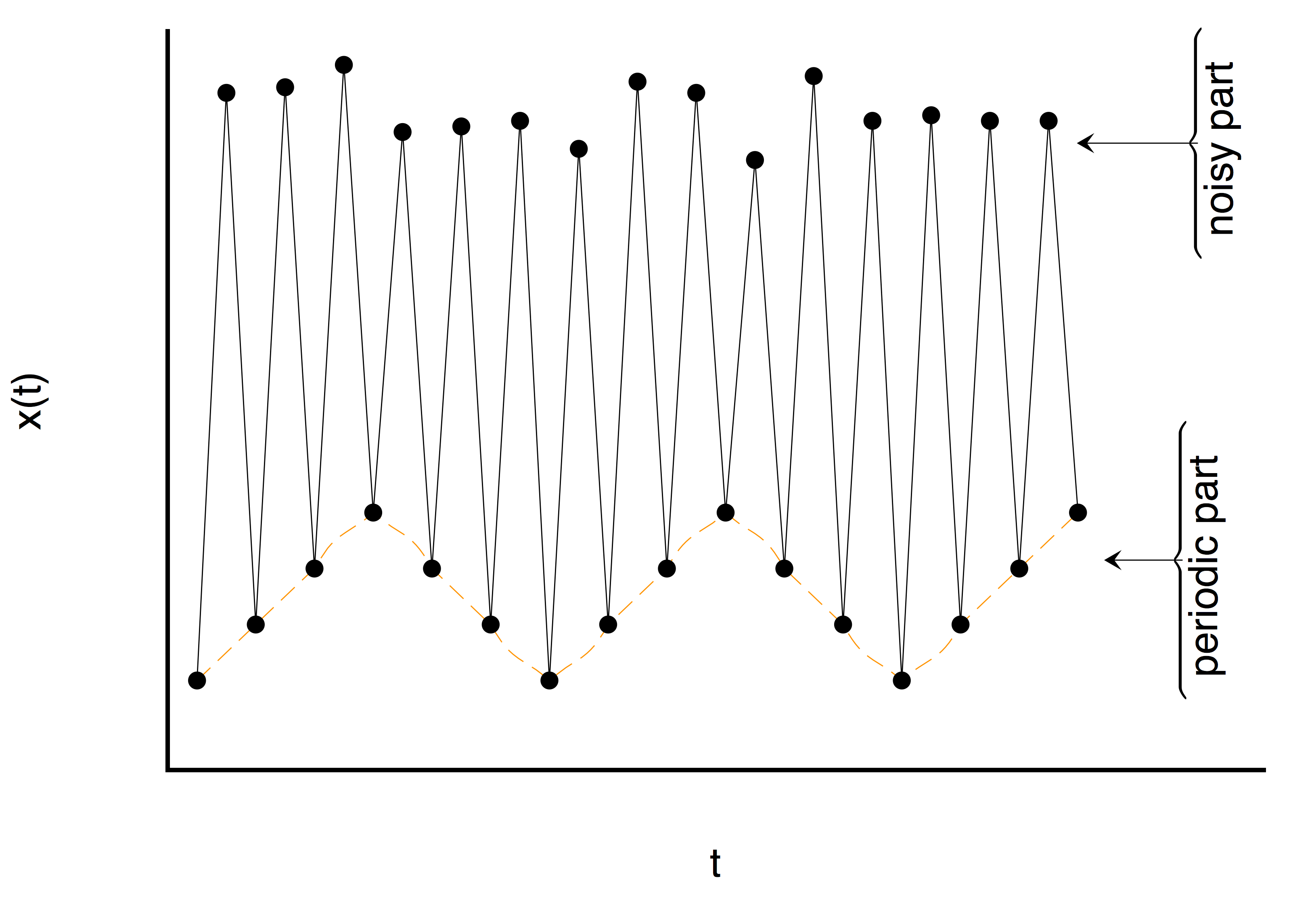}
\caption{Cartoon series having a noisy part at even timesteps and a periodic progression part at odd timesteps: a classical VG/HVG analysis fails to capture the hidden periodic pattern, whereas such pattern can be easily extracted from $G_{\text{bot}}$.}
\label{fig:better}
\end{figure}

\noindent Here we label as $G_{\text{top}}(x(t))$ (i.e. the `top' visibility graph) to the graph extracted for either of the two aforementioned procedures\footnote{at this point we consider both kinds of graphs VG and HVG indistinctively, although we know that they indeed provide different sorts of information and in practice depending on the particular processes under study it will be more adequate to make use of either HVG or VG}. The label 'top' comes from the fact that visibility is indeed applied 'from above' and therefore tends to encapsulate information on the relative position of local maxima (peaks, see figure \ref{fig:VGHVG} for an illustration).  An obvious drawback of a basic VG/HVG representation is that local minima are hidden, and more specifically the elements in the {\it pit} set defined above by construction map into nodes with a fixed degree $k=2$, independently of their actual value.  Consider for instance a signal whose odd values build a periodic progression, and whose even values are just noise (see Fig. \ref{fig:better} for an illustration). In this simple example, the periodic structure is completely hidden if one only looks at the standard VG/HVG: structure on the {\it pit} set is lost. As a result, VG/HVG might be insensitive to processes which have two or more spatial scales\footnote{For another simple example, consider two time series, the first being a periodic series of period 2, and the second being a mix of two processes: uniform white noise in $[0,1]$ for even times $t=2p, p\in \mathbb{N}$, and a constant value $>1$ for odd times $t=2p+1, p$. The associated VG/HVG is identical despite that both processes are qualitatively different, however using both top and bottom graphs the difference is obvious.} However, this drawback is removed if the VG/HVG algorithm is additionally and subsequently applied  'from below'. Accordingly, one can also define a 'bottom' visibility graph $G_{\text{bot}}(x(t))$ where the visibility criterion is now applied from below, which now will focus {\it particularly} on the structure of the local minima (highlighting in particular the connectivity structure of the {\it pit} set) as recently observed \cite{previous}. Note that this procedure is performed on the whole signal, hence when constructing $G_{\text{top}}$ and $G_{\text{bot}}$ one is not discarding information on the intermediate data (as it happens with traditional time-domain approaches described above). Furthermore, it is easy to prove that the bottom construction coincides with the top construction if applied on the flipped series, in such a way that the following identities hold: 
$$\forall {\cal X}: G_{\text{bot}}(x(t))=G_{\text{top}}(-x(t)), \ G_{\text{bot}}(-x(t))=G_{\text{top}}(x(t)).$$ 
\noindent Our working hypothesis therefore exploits the potential differences between the top and bottom graphs as a proxy for quantifying the difference between local maxima and local minima statistics in $\cal X$, or the {\it top-bottom VG/HVG asymmetry} ($\Delta$VGA) for short. Mathematically, $\Delta$VGA can be quantified in very many different ways (different protocols can be defined in an {\it ad hoc} way depending on the particular problem under study). For instance, simple {\it global} topological features that we know are informative include statistics of the degree sequence \cite{Luque2016} such as graph's degree distribution $P(k)$, for which, adopting the ${\ell}_1$ norm distance between distributions, leads to a particular definition of the asymmetry $\Delta\text{VGA}=\sum_k |P^\text{top}(k)-P^\text{bot}(k)|$. This is just an informed choice and in any event, it always comes down to a comparison between certain set of features extracted from the top and bottom VG/HVG.\\
%but in general the scheme systematically proceeds by extracting certain topological features $f_i$ from the top and bottom graphs and define distances over these $d_i=||f_i^{\text{top}}-f_i^{\text{bot}}||.$
{As mentioned above, there are a number of interesting scenarios involving differences in the marginal distributions or correlation structure of peaks and pits (${\cal S}_1$, ${\cal S}_2$, ${\cal S}_3$), and $\Delta$VGA can serve as a tool to investigate these asymmetries. While these issues have not been fully examined in prior work, one important study \cite{previous} has used a comparison between $G_{\text{top}}$ and $G_{\text{bot}}$ with the intention of studying the different dynamics of local minima and maxima in the particular case of sunspot time series. As a matter of fact, sunspot series can be considered the degenerate case because they have a natural absolute and frequently occurring natural minimum (zero), which \textit{by definition} imposes different features on the local minima  (including impacting their variance, serial autocorrelation and power spectra). Similar cases where dampening function is applied ($ \text{if}\  x > y \Rightarrow x:=y$) are similarly unhelpful.} \\

\noindent Before presenting the practical algorithmic protocols we explore in synthethic processes the performance of this method, and in particular the ability to outperform current methods as well as the transversality (e.g. how well scenarios ${\cal S}_{1-3}$ above are addressed).
\begin{figure}
\centering
\includegraphics[width= 14 cm]{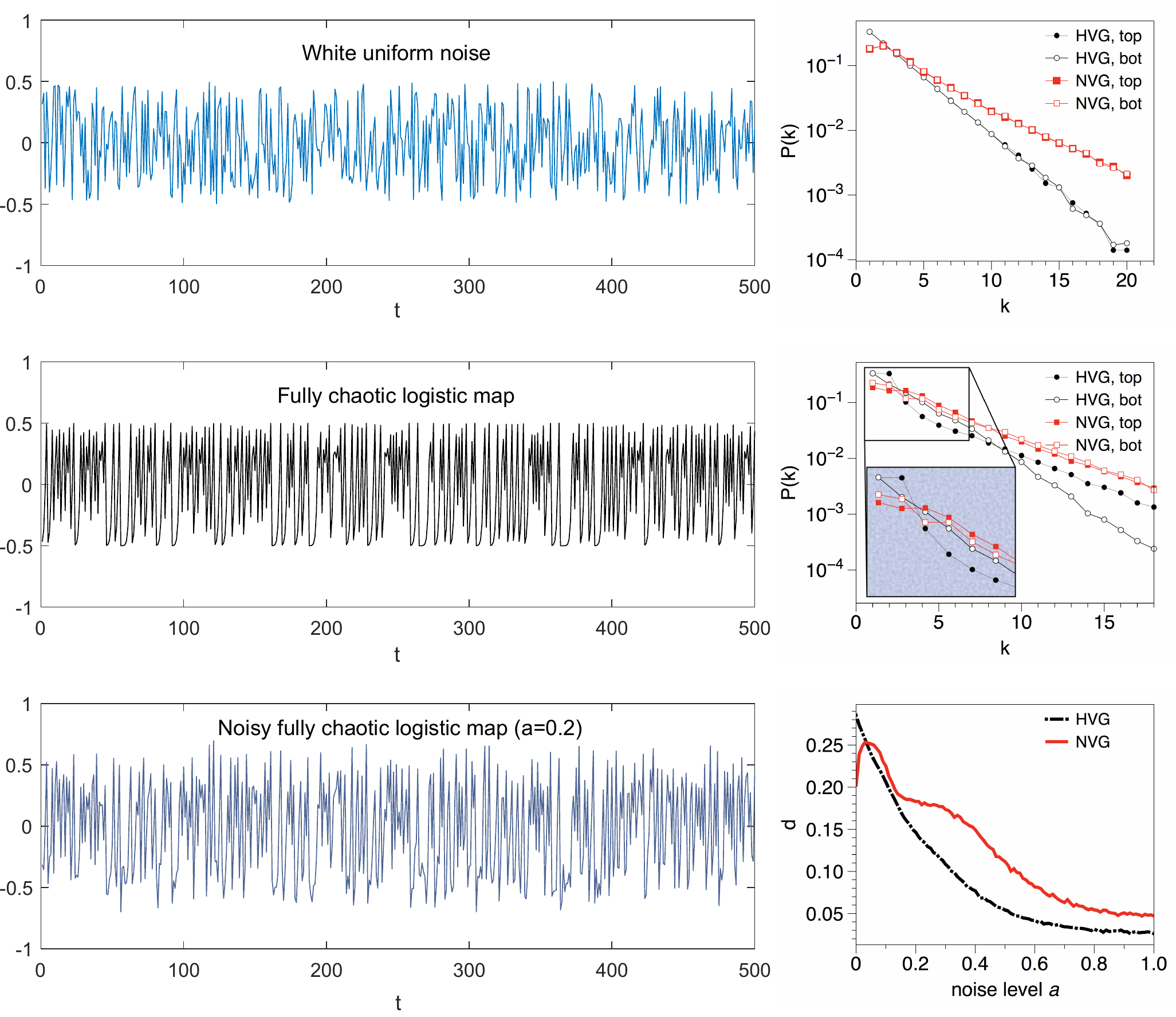}
\caption{(Left panels) Sample time series of uniform white noise (top panel), fully chaotic logistic map (middle panel) and fully chaotic logistic map polluted with a certain amount of white noise (bottom panel). In all cases the signals are irregular and lack any obvious pattern (for appropriate visual comparison all signals have been scaled in the vertical axis). However, whereas the statistics of $G_{\text{top}}$ and $G_{\text{bot}}$ are equivalent for i.i.d., this is not the case for the chaotic process. In the last case, noise pollutes and hides the chaotic signal and as such the statistics of $G_{\text{top}}$ and $G_{\text{bot}}$ depend on the noise amplitude. (Right panels) Comparison of the degree distributions extracted from $G_{\text{top}}$ and $G_{\text{bot}}$ (for both VG/HVG). For the uniform white noise, distributions coincide as expected (the process is invariant under $x(t)\to -x(t)$). For the fully chaotic logistic series, statistics of $G_{\text{top}}$ and $G_{\text{bot}}$ are different and this feature evidences differences between local minima and maxima. In the last panel we plot the distance between distributions (using $\ell_1$ norm) for a fully chaotic logistic map polluted with a certain amount of noise (see the text).}
\label{fig:samples}
\end{figure}

\subsection{\textsf{Theoretical validation on synthetic processes}}
To validate the method, we initially consider a battery of dynamical processes with varying degrees of complexity which focus on different aspects, and consider in every case the performance of a particular definition of $\Delta\text{VGA}=\sum_k |P^\text{top}(k)-P^\text{bot}(k)|$ (measured on both VG and HVG) and their comparison with the AVA-based metric (VR) (see figure \ref{fig:samples} for an illustration). Results are summarized in table \ref{table:2}. We can highlight the following key results: \\
\begin{enumerate}
\item Symmetric stochastic processes (both with no correlations --white noise-- and with rapidly decaying correlations --red noise-- yield statistically identical top/bottom VG/HVGs and thus vanishing values of $\Delta$VGA. 
\item For non-symmetric white noise, peak and pit statistics are different, even if the process lacks information. In that case, both $\Delta$VGA (applied to VG) and VR detects such asymmetry, while $\Delta$VGA applied to HVG filters out dependences solely based on marginals and predicts no difference: in this sense we conclude that HVG does not capture scenario ${\cal S}_1$ as defined above.\\

\noindent These two initial observations can be summarised in the following theorem:\\

\noindent {\bf Theorem 1.} {\it Let $\{x(t)\}$ be a bi-infinite time series generated by (i) white noise with a continuous probability density $f(x)$ or by (ii) symmetric correlated red noise. Then $\Delta$\text{VGA}$=0$ for HVG.}\\
A proof for this theorem can be found in the appendix.
\item For chaotic processes  where peaks and pits are different both in terms of marginal statistics (scenario ${\cal S}_1$) as well as in terms of correlation dependences (scenarios ${\cal S}_2$ and ${\cal S}_3$), all methods successfully detect such asymmetry. In particular, the following theorem holds:\\

\noindent {\bf Theorem 2.} {\it Let $\{x(t)\}$ be a bi-infinite time series generated by a fully chaotic logistic map. Then $\Delta$\text{VGA}$>0$ for HVG.}\\
A proof for this theorem can be found in the appendix. Gathering together theorems 1 and 2, we have rigorously proved that a chaotic process such as the fully chaotic logistic map can be easily distinguished from white noise, despite the fact that both processes have a flat power spectrum (delta-distributed autocorrelation function). Additionally, we numerically observe that chaotic processes with a fast-decaying correlation structure are also distinguishable from exponentially decaying correlated noise.
\item Interestingly, there are several cases (such as processes with two alternating dynamics for peaks and pits) where currently used indicators (AVA, AFA) will fail, while $\Delta$VGA efficiently captures significant differences (i.e. VR does not capture scenarios ${\cal S}_2$ and ${\cal S}_3$ as defined above). 
\item $\Delta$VGA is robust against noise pollution and works for short time series, enabling its use in practical cases.
\item  VR is a quantity which by construction only depends on the marginal distribution of peaks and pits (in particular, the variance of these marginals), and as such it does not carry information on any temporal correlations, neither intra peaks or intra pits (scenario ${\cal S}_2$), nor inter peaks/pits (scenario ${\cal S}_3$). It is easy to prove that for a given series $x(t)$, if one breaks down all temporal correlations by reshuffling the series, then the new, reshuffled series $x^{\text{reshuffled}}(t)$ and $x(t)$ will still have the same marginal distributions and thus (under some assumptions) the same VR value, yet $x^{\text{reshuffled}}(t)$ is indeed white noise with a flat spectrum and delta-distributed autocorrelation function, very different in general from the non-reshuffled series $x(t)$. In the same line, VR {\it breaks down} for any signal which is composed by two alternating processes with similar variances and dynamics with different invariance properties, whereas $\Delta$VGA is not afflicted by these drawbacks.\\
\end{enumerate}

\begin{figure*}
\includegraphics[width=0.6\columnwidth]{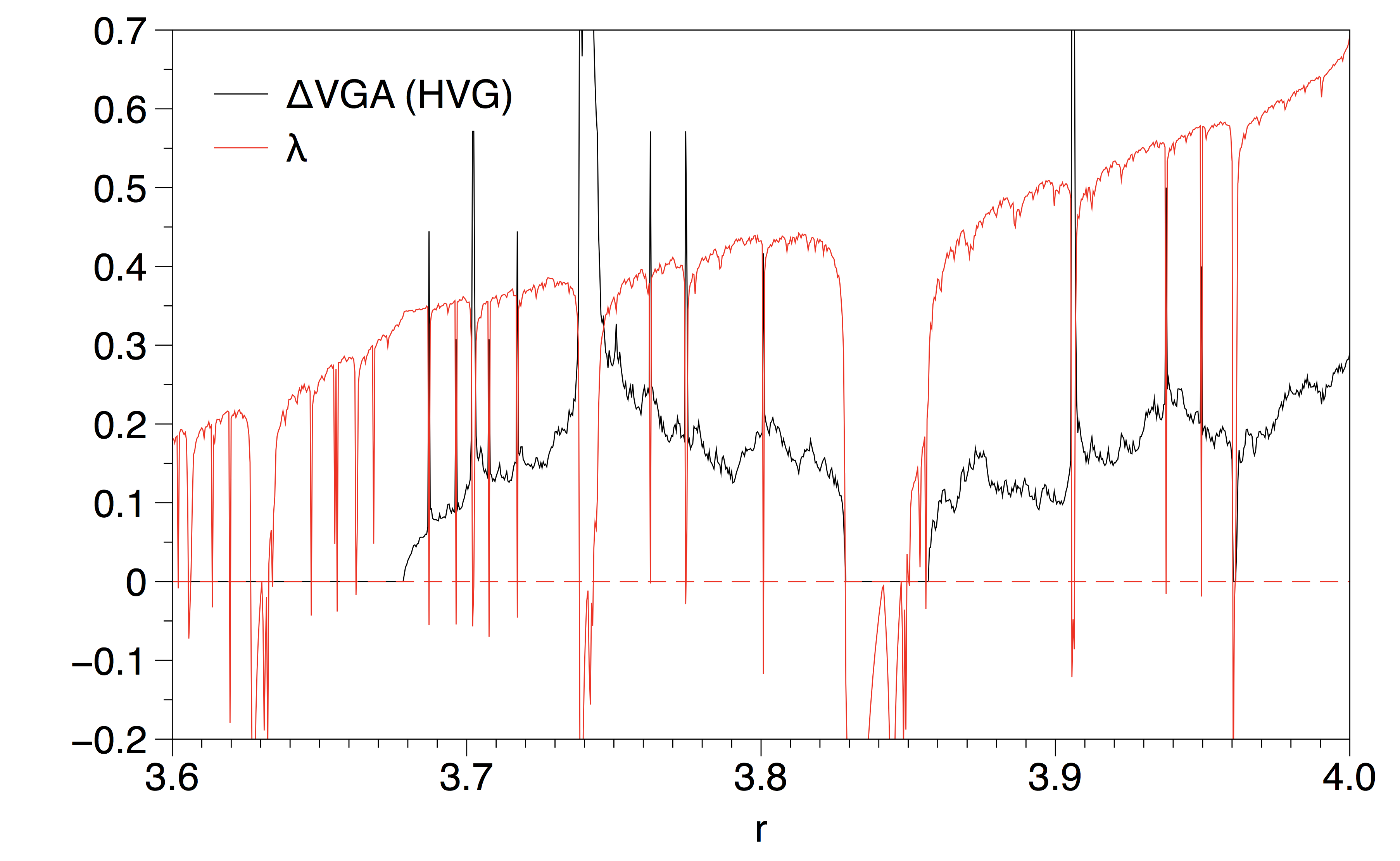}
\caption{Plot of $\Delta$VGA (defined as the $\ell_1$ distance between top and bottom degree distributions) for time series extracted from the logistic map $x(t+1)=\mu x(t)[1-x(t)]$, for a range of values of $\mu$ for which the map generates chaotic trajectories of different type, interspersed by periodic windows. In the same figure we plot the Lyapunov exponent of the map.}
\label{parametric}
\end{figure*}

\subsection{\textsf{Parametric analysis}}
\noindent Finally, in order to explore the relationship between $\Delta$VGA and a number of standard linear and nonlinear indicators (mean, variance, skewness, kurtosis, Lyapunov exponent), we have considered a parametric map (the logistic map) $x(t+1)=\mu x(t)[1-x(t)]$. As $\mu$ smoothly varies between $3.6$ and $4$ the map generates chaotic (aperiodic) time series with different degree of chaoticity (different Lyapunov exponent, attractor's fractal dimension and physical invariant measure), interspersed by periodic windows. We recorded each statistic as we continuously scan $\mu$. 
In figure \ref{parametric} we plot $\Delta$VGA (defined as the $\ell_1$ distance between top and bottom degree distributions for time series) extracted from the logistic map $x(t+1)=\mu x(t)[1-x(t)]$, for a range of values of $\mu$ for which the map generates chaotic trajectories of different type, interspersed by periodic windows. In the same figure we plot the Lyapunov exponent of the map. Both indicators are only weakly correlated (Pearson correlation coefficient $r=0.22$). Correlation between $\Delta$VGA and other standard indicators include: $r=-0.104$ (mean), $r=0.2$ (standard deviation), $r=-0.12$ (skewness), $r=0.158$ (kurtosis). We conclude that $\Delta$VGA finds negligible correlation with all indicators, except for standard deviation and Lyapunov exponent, where the correlation was found to be weak:  $\Delta$VGA provides genuinely different information than standard linear and nonlinear indicators. Additionally, note that $\Delta VGA$ is notably unrelated to power spectrum (for instance, $\Delta$VGA is positive for $\mu=4$ and null for white noise, and both processes have identical power spectrum).\\

\begin{table}
\centering
%\begin{ruledtabular}{\columnwidth}
 \begin{tabularx}{\linewidth}{|X|X|X|cc|c|}
%\begin{tabular}{|c|c|c|ccc|}
\hline
&&&$\Delta$VGA& \text{of :}&\\
{\bf Process}& {\bf Definition}&{\bf Expected outcome}&VG&HVG&AVA\\
 \hline
 (symmetric) uniform white noise & $x(t)=\xi$, $\xi \sim U[-1,1]$ & 0 & \checkmark & \checkmark & \checkmark \\
 \hline 
exponentially correlated red noise&$x_t=r x_{t-1}+\xi, \quad \xi \in N(0,1)$&0 &\checkmark&\checkmark&\checkmark\\ 
  \hline
 Power law (non-symmetric) white noise&$x(t)=\xi$, $\xi \sim x^{-\alpha}$&0&\checkmark& $\times$ &$\times$ \\
  \hline
 Fully chaotic logistic map & $x_t=4x_{t-1}(1-x_{t-1})$ & $>0$ & \checkmark&\checkmark&\checkmark\\
  \hline
 Fully chaotic logistic map polluted with white noise & $\rho_a(t)=x(t)+\xi$, $\xi(t)\sim  \mathcal{U}(-a,a)$,  $a \in [0,1]$, $x(t)$ fully chaotic logistic map. & $>0$ but decreasing with noise amplitude & \checkmark & \checkmark & \checkmark \\
  \hline
Selected pollution  & 
$y_t=x_t+\xi$,  $x_t$ fully chaotic logistic map, 
$\xi=\begin{cases} \sim U[0,a], \ x_t<0.3\nonumber\\
                   0 \quad \text{otherwise}
     \end{cases}$,

 & $>0$ but decreasing with noise amplitude & \checkmark & \checkmark & \checkmark \\
  \hline
 Alternating dynamics $\#1$: $x_t$ fully chaotic logistic, $\xi \sim U[1-a,a]$ & 
$y_t=\begin{cases} \sqrt{8}ax_t,\ t\ \text{even}\\
                   \sqrt{12}\xi \quad \text{otherwise}
                  \end{cases}$                                  
 & $>0$ & \checkmark & \checkmark & $\times$ \\
 \hline 
  Alternating dynamics $\#2$: $x_t$ fully chaotic logistic, $\xi \sim U[1-a,a]$&
$Y_t=\begin{cases} Ca\cdot \sin(\pi t/20),\quad t\ \text{even}\\
\sqrt{12}\xi \quad \text{otherwise}
\end{cases}$ where $C$ is such that the variance of the sinusoidal process is one                      
 & $>0$ and independent of $a$ until both processes coalesce ($a>0.5$) & \checkmark & \checkmark & $\times$ \\
 \hline
\end{tabularx}
%\end{ruledtabular}
\caption{ \label{table:2} Summary of results on synthetic processes. In every case we extract time series from the dynamical equations defined on the first column, and compare the results provided by the statistics $\Delta$VGA (applied to VG and HVG) and AVA (computed from the logarithm of the Variance Ratio), see the text for details. }
\end{table}

\noindent From the theoretical analysis above, we conclude that $\Delta$VGA as defined above can be efficiently used to detect and quantify subtle differences between the statistics of local maxima and local minima in the associated series which are not correlated with standard linear and nonlinear indicators.  If such asymmetry is solely based on the (one-point) marginal distributions of peaks and pits (scenario ${\cal S}_1$), then all three choices ($\Delta$VGA applied to VG and HVG, and AVA) are qualitatively similar. However, in the event the peaks and pits have identical marginals but different correlation structure, then only $\Delta$VGA capture this difference. Finally, if there is a difference between marginals but no difference in the correlation structure, then $\Delta$VGA (applied to VG) and AVA accurately capture the difference, but not $\Delta$VGA applied to HVG (as this latter is an order statistic).\\

\section{\textsf{Data extraction methods and visibility protocol}}
We have applied the methodology in three different situations: EEG-fMRI data, financial time series of NYSE, and worldwide temperature records. In what follows we provide details on data acquisition and pre-processing, as well as the protocols devised in each case to compute $\Delta$VGA. 
\subsection{\textsf{EEG-fMRI data}}
\subsubsection{Data acquisition and artifact correction}
Data was collected during a synchronous EEG-fMRI acquisition protocol, which was approved by the Ethical board of Goethe University (Kommission des Fachbereichs Medizin der J. W. Goethe-Universitat Frankfurt am Main as of January 10th, 2008). This data has been analyzed in several prior studies that examined: identification of K-complex correlates in N2 sleep \cite{Jahnke12}, use of functional connectivity for sleep staging \cite{Tagliazucchi12}, impact of sleep on serial autocorrelation of BOLD time series \cite{Tagliazucchi13}, and our evaluation of AVA \cite{Davis16}. Here we only used the EEG data to determine sleep stages. EEG was sampled at an initial sampling rate of 5 kHz, low pass filtered at 1 kHz, and down-sampled to 250 Hz for artifact cleaning and sleep staging and further analysis (see below). Data were recorded using a BrainCapMR (Easycap) EEG cap (Herrsching, Germany) with 30 recording channels. The MR-compatible amplifiers were BrainAmp MR+, BrainAmp ExG; BrainProducts (Gilching, Germany). Data were recorded using BrainProduct's ``Recorder" and analyzed using Brain Product's ``Analyzer". Analysis steps included: MRI and pulse artifact correction performed based on the Average Artifact Subtraction method \cite{Allen98} as implemented in Vision Analyzer2 (Brain Products, Germany) followed by objective (CBC parameters, Vision Analyzer) ICA-based rejection of residual artifact-laden components after artifact subtraction. Good quality EEG was obtained, which allowed for sleep staging by an expert, according to criteria of the American Academy of Sleep Medicine \cite{Berry12}. Sleep staging was based on scoring 30s blocks of the EEG data.  Based on this scoring we ignored sections associated with transitions between sleep stages, maintaining  only those without transitions. BOLD time series sections matching these were spliced from the recorded time series.
Functional MRI scans were acquired on a 3T system (Siemens Trio, Erlangen, Germany) using single-shot T2*-weighted EPI (32 slices, repetition time/echo time $= 2,080 ms/ 30 ms$, matrix $= 64 \times 64$, voxel size $= 3 \times 3 \times 2 mm^3$, distance factor $= 50\%$). To correct for physiological noise, physiological responses (cardiac, respiratory) were recorded through sensors from the MR scanner (sampling rate $= 50 Hz$) and MR-compatible devices (BrainAmp MR+, BrainAmp ExG; Brain Products). Sixty-three healthy non-sleep-deprived participants (thirty-six females, mean $\pm$ SD age of $23.4\pm3.3$ years) were scanned in the evening (starting from 8 PM). Data from those 55 participants who reached at least sleep stage N1 was used in our analysis. 
\subsubsection{\textsf{fMRI epoch selection and pre-processing}}
Each of the 55 participants provided epochs of functional imaging data during wakefulness (W) and at least the N1 sleep stage. From these epochs, we set the minimal time series length at 135.2 seconds (65 volumes), amounting to 293 total epochs. In practice, most of the epochs were 5-10 times as long (see \cite{Davis16} for details and quality control procedures). 
\begin{comment}
IN CASE WE WANT TO MAINTAIN QC PROCEDURES
We conducted a quality control procedure in which we examined the functional time series for loss of signal in any volume, and derived temporal SNR histograms for each epoch to identify whether any epoch was associated with a shifted distribution indicating low quality data. Authors B.D and U.H conducted this procedure jointly. 
\end{comment}
We used AFNI \cite{Cox96} for pre-processing and physiological-noise correction (PN-correction). De-spiking of the time series was carried out as part of the PN-correction workflow using AFNI's 3dDespike. This is important, because the procedure replaces time series that exceed 2.5Standard deviations of the mean with values estimated by smoothing to nearby values. This reduces the relative impact of extreme values. 
The physiological (cardiac, respiratory) data were down-sampled to the acquisition rate of single volume slices (15.4Hz). We used AFNI's \textit{retroTS.m} procedure to create slice-based regressors from these data. This utility generates, for each physiological regressor, a time series that is phase shifted to match the timing of each slice's acquisition. From the cardiac and respiration recordings we derived 13 such slice-based regressors: 4 for the cardiac series and its harmonics, 4 for the respiratory series and its harmonics, and 5 for respiration variation over time and its harmonics, \cite{Birn06}. We removed the variance explained by these 13 regressors from the BOLD time series using linear regression (RETROICOR; \cite{Glover00}) as implemented in AFNI.\\

\noindent Following PN-correction, we discarded the first 5 volumes of each epoch from the analysis to allow for T1 stabilization effects, and then performed slice timing correction (3dTshift), motion correction (3dvolreg), and spatial smoothing (3dmerge, 6mm FWHM Gaussian Kernel) on all of the images. We then removed several sources of variance from the time series data via linear regression. These included (i) 6 motion parameters estimated during the head motion correction, and (ii) linear, second-order and third-order polynomial trends. We used the residuals of the regression procedure as the functional MRI data of interest in all subsequent analyses.
\begin{comment}
NOT SURE WE NEED THIS IN SCOPE OF THIS PAPER
Because we partialed out physiological-related variance from direct measurements we did not use proxy measures typically used for this purpose such as data from CSF voxels or White Matter voxel. While these are potential proxies for physiological time series, they are only moderately correlated; see Chang \& Glover, 2009). Furthermore, given the impact of global-mean correction on functional connectivity estimates (see Murphy et al., 2009; Schölvinck et al., 2010), we did not implement this procedure.\\ 
\end{comment}

\begin{comment}
UH: I THINK THE NEW FIGURE DEALS WITH THIS WELL
\begin{figure}
\includegraphics[width=12 cm]{POC.png}
\caption{\textcolor{red}{This is basically an illustration of the method, similar as figure \ref{fig:VGHVG}, so we need to produce a good figure (and only one)}}
\label{fig:fmri1}
\end{figure}
\end{comment}

\subsection{\textsf{Calculating $\Delta$VGA and protocols}}

\subsubsection{Neurobiological case}
In a neurobiological context we were interested in monitoring systematic differences between regions of the degree distributions (rather than a net value out of the comparison of the whole distributions) which hold across participants in specific brain areas. A scalar projection such as the one defined for $\Delta$VGA in the synthetic cases lose such fine-grained level of detail. Therefore, in this particular application we first constructed, for each voxel, degree distributions up to degree $k=9$ (this histogram truncation is justified as the modal degree value in fMRI series is around 3 or 4 with very few time points having a degree greater than 8, as shown in the results section). A second reason for selecting $k=10$ as a limit was our interest in local features of the time series, rather than in any high-amplitude but infrequent spikes isolated from each other by more than 22 (TR = $2.2s \times 10$ samples) seconds. We recall that Figure \ref{fig:VGHVG} illustrates graphically the concept of how node degrees are established, and how a time series can be characterized via degree distributions of the top and bottom VG/HVGs. As an additional pre-processing, in this application we created empirical cumulative distribution (CDF) functions of each histogram, normalizing the bar heights so that the area of the histogram is equal to one.  \\  

%\subsection{\textsf{Group Analysis}}
\noindent {\bf Normalization to common spatial reference space. }After creating maps of visibility graph differences (top-bottom)  on a single voxel level, we obtained a transformation between each EPI time series to its corresponding anatomical image using FSL's epi-reg script. The most important steps in this procedure are FAST's (FMRIB's Automated Segmentation Tool; \cite{Zhang01} histogram based segmentation of the T1 structural scans to derive white matter maps, and the use of the boundaries of these white matter maps to perform Boundary-Based co-Registration of the EPIs to their corresponding T1 structural images (BBR; \cite{Greve09}). We then performed nonlinear normalization (FNIRT), of each subject's T1 images into $2\times 2 \times2$ MNI space. Importantly, we concatenated the two transformations (EPI to T1; Linear, guided by white matter boundaries) and T1 to MNI (nonlinear warp) to derive a single transformation from EPI space to MNI space. We used this transformation to align the AVA maps from original space to the MNI template in a single step.\\

\noindent {\bf Single-voxel calculations, and group-level analyses. } The sleep-staging procedure allowed us to obtain fMRI time series for wakefulness (W) and three different sleep stages: N1, N2 and N3. On the single participant level, for each of the four conditions, we conducted voxel-wise  analyses to derive the node-degree histograms for the top and bottom graphs (i.e. these were derived for each voxel). These were represented as the empirical cumulative distribution function.
% and refer to these as WTOP, WBOTTOM, N1TOP, N1BOTTOM, N2TOP, N2BOTTOM, N3TOP, and N3BOTTOM. 
With these CDFs we could answer the following two questions: 
\begin{enumerate}\item	On the single-condition level we identified brain areas that differentiated the top from bottom CDFs; we refer to this `difference' histogram as {\it Asymmetry of Visibility Histogram} (AoVH), which is our primary way of implementing $\Delta$VGA in this application. We performed these analyses for CDFs derived from both VG and HVG. It is important to note that because these CDFs were normalized, a main effect of condition is not possible (the histograms average across bins to one) and there will always be a strong main effect of bins. Instead, it is the interaction between top/bottom analysis and Bin which we use to show differences across conditions. In reference to the illustration in Figure \ref{fig:VHGVG} (panel C), we are interested in whether the difference between the blue and red histograms, calculated for a given voxel's time series' are systematic across participants.
% * <lucas_lacasa@yahoo.es> 2017-06-29T18:15:44.089Z:
%
% ^.
\item Then, in order to study differences between conditions, we compared the AoVH of the two conditions. We did this by deriving AoVH for each condition and then determining statistically whether these differed between the two conditions.  Note that comparisons between conditions were based on time series matched for length within participant.
\end{enumerate}

\noindent On the group level, to evaluate voxels showing statistically-significant AoVH \emph{within} each study condition (W, N1, N2, N3) we conducted voxel-wise repeated measures ANOVA with two fixed factors; Time series view (two levels [Top, Bot]), and node-degree-histogram Bin (8 levels as explained above). To enable inferences about the population, we modeled participants as a random factor. In order to compare \emph{between} any two conditions (e.g., W vs. N1) we derived the AoVH for each condition, and conducted a repeated measures ANOVA with 2 fixed factors: Condition (e.g., W vs. N1) and Node degree histogram bin of the difference histogram (8 bins, values 2(min)-9(max)). We note that 8 bins are used here, as the 9th bin would not be independent of the prior 8 (due to normalization).\\ 

These analyses returned a statistical significance value for the interaction term, for each brain voxel. We then implemented a clustering procedure \cite{forman99} to identify brain areas where many contiguous voxels, each with a $p$ value of $\leq$ .001 show a significant interaction: this identifies an `activation cluster'.\\

\subsubsection{Financial case}
In application to financial data, we considered a dataset of financial stocks comprising stock evolution of 35 major American companies from the New York Stock Exchange (NYSE) and NASDAQ in the period 1998-2012, the majority of which belong to the Dow Jones Industrial Average. Data have been extracted from \cite{Flanagan2016}.\\
In previous sections we have confirmed that the ($\ell_1$) distance between  top and bottom degree distributions of VG/HVG  is an efficient and informative definition of $\Delta$VGA which in many cases outperforms state of the art signal peak/pit asymmetries and is genuinely different from standard linear and nonlinear indicators. This is precisely the scalar which we shall use in the financial context. Our protocol is as follows: we compute the $\Delta$VGA by splitting the time series for each company into yearly time series and calculate the distance $d=\sum_k |P^\text{top}(k)-P^\text{bot}(k)|$ over the associated HVGs for each year. Accordingly, each year is characterized by a vector of distances. For instance, for year 1998, we have a vector of 35 dimensions whose entry $i$ is the $\Delta$VGA for company $i$ in 1998. Subsequently, a principal component analysis is performed to dimensionally reduce data, and a projection into the two first principal components is used to visually cluster different years.

\subsubsection{Climatic case}
This application is conceptually similar to that reported for financial time series. Rather than considering stock prices of major shares in multiple years, we considered daily temperature data sampled over a global grid with a resolution of 192 longitude and 94 latitude, with average temperature for each day (365 values) provided for each grid point (data obtained from \cite{Kanamitsu2002}). For each year between 1995 and 2015 we compute $\Delta$VGA following the same definition used in the financial setting.

\section{\textsf{Results}}

\subsection{\textsf{Application to fMRI}}
\begin{figure}
\includegraphics[width=12 cm]{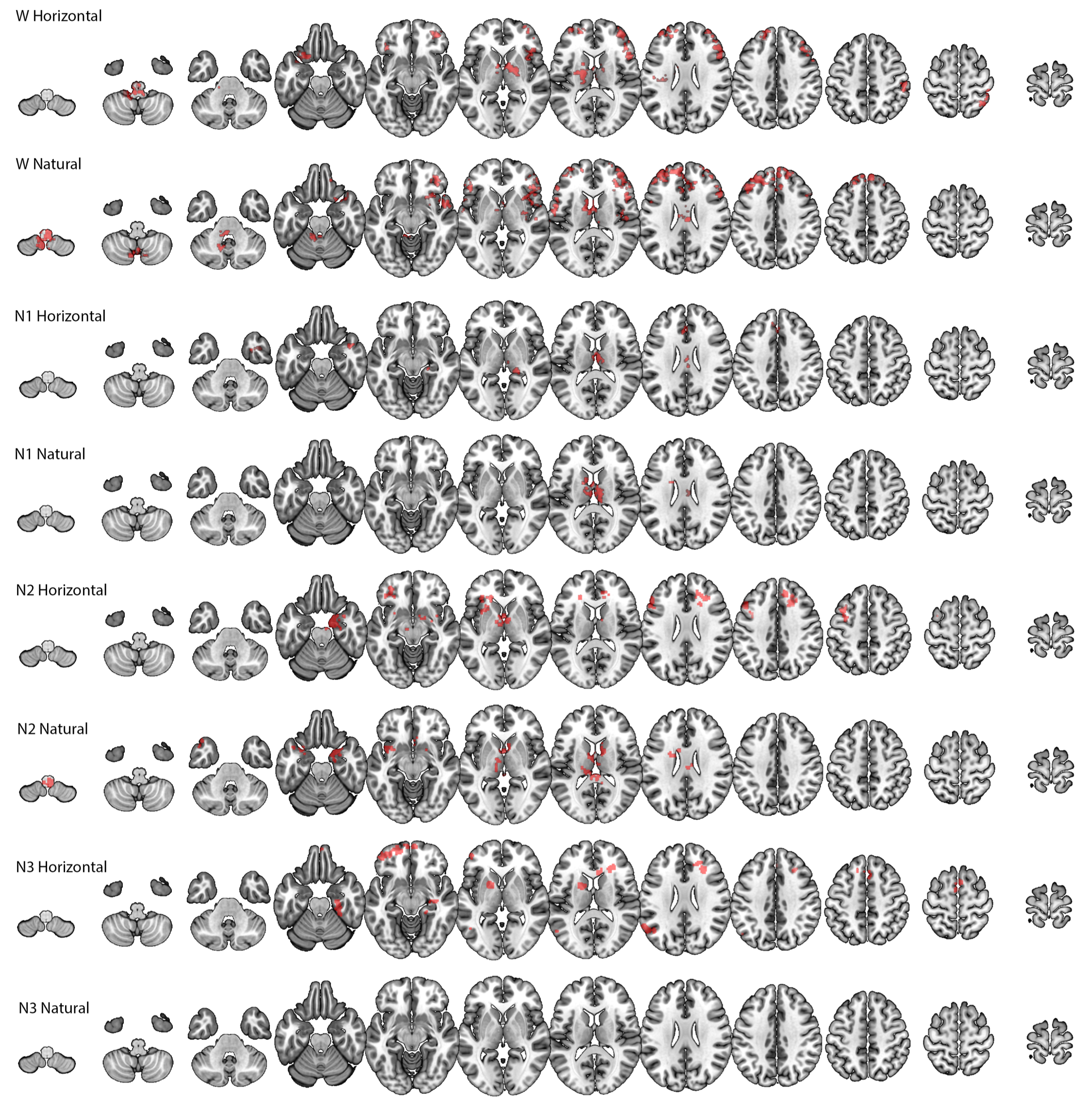}
\caption{For wakefulness (W) and each sleep stage (N1, N2, N3), the figure shows brain regions for which the degree distributions of the top and bottom graphs differed significantly as determined by an ANOVA repeated-measures analysis across participants. Horizontal=derivation from HVG; Natural=derivation from VG.}
\label{fig:BRAIN}
\end{figure}

We first compared the top and bottom visibility graphs \emph{within} the Wakefulness and the three sleep stages (N1, N2 and N3). As shown in Figure \ref{fig:BRAIN}, we found significant differences, predominantly in thalamic and frontal regions, for each of these  conditions, with the exception of the case of VG in the N3 sleep condition.\\ 
\noindent To better understand these results, we determined which visibility values tended to contribute most strongly to the statistically-significant differences in degree distributions that produced the Wakefulness (W) results in Figure \ref{fig:BRAIN} (rows 1, 2). To this end, for each cluster we derived a histogram that communicated the visibility bins that most strongly differentiated the top and down degree distributions for each voxel in the cluster. We did this by (i) transforming the cluster to original space, (ii) for each voxel, identifying the bin that maximally differentiated the top from down histogram, (iii) transforming that value to common space, (iv) creating an average across participants for each voxel in the cluster.  The resulting histograms communicated a very clear and consistent result: for VG the modal degree value that maximally differentiated the top and bottom histograms was 4, with narrow tails towards the values of 3 and 5 (indicating that for some participants, some voxels maximally differentiated the histograms at values 3 and 5). Importantly, there were no cases with means below 3 or above 5. For HVGs, in \emph{all} clusters the modal degree value that maximally differentiated the top and bottom histograms was 3, with very narrow tails towards 2 and 4.\\
These findings are very important as they show that the differences identified by $\Delta$VGA (AoVH) were driven by very local dynamics rather than due to differences in propensity of isolated extreme events.\\

\begin{figure}
\includegraphics[width=12 cm]{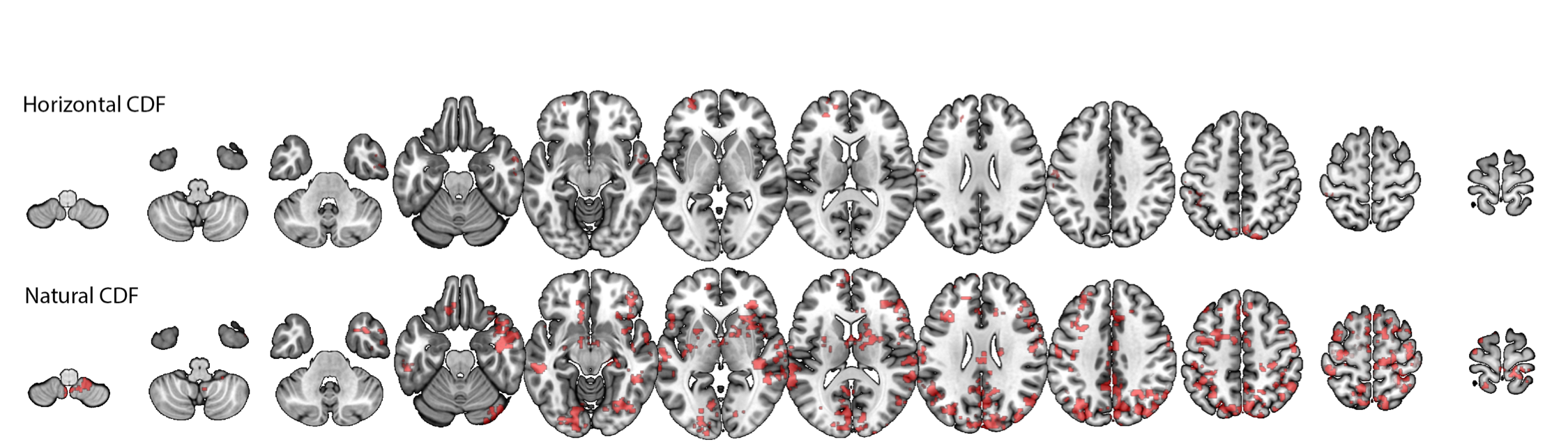}
\caption{Brain areas where spontaneous activity patterns differed between wakefulness and N2 sleep. In these areas, the difference between top and bottom graphs varied systematically between wakefulness and sleep.}
\label{fig:BRAIN2}
\end{figure}

\noindent In a separate analysis, we also found that $\Delta$VGA profiles could discriminate wakefulness from sleep. As shown in Fig.\ref{fig:BRAIN2}, we identified numerous areas, in both occipital (visual cortex) and lateral temporal cortices, where dynamics during wakefulness and N2 sleep differed significantly. No differences were found between W and N1 or between W and N3.

\subsection{\textsf{Application: unsupervised clustering of financial periods}}
\begin{figure*}
\includegraphics[width=0.43\columnwidth]{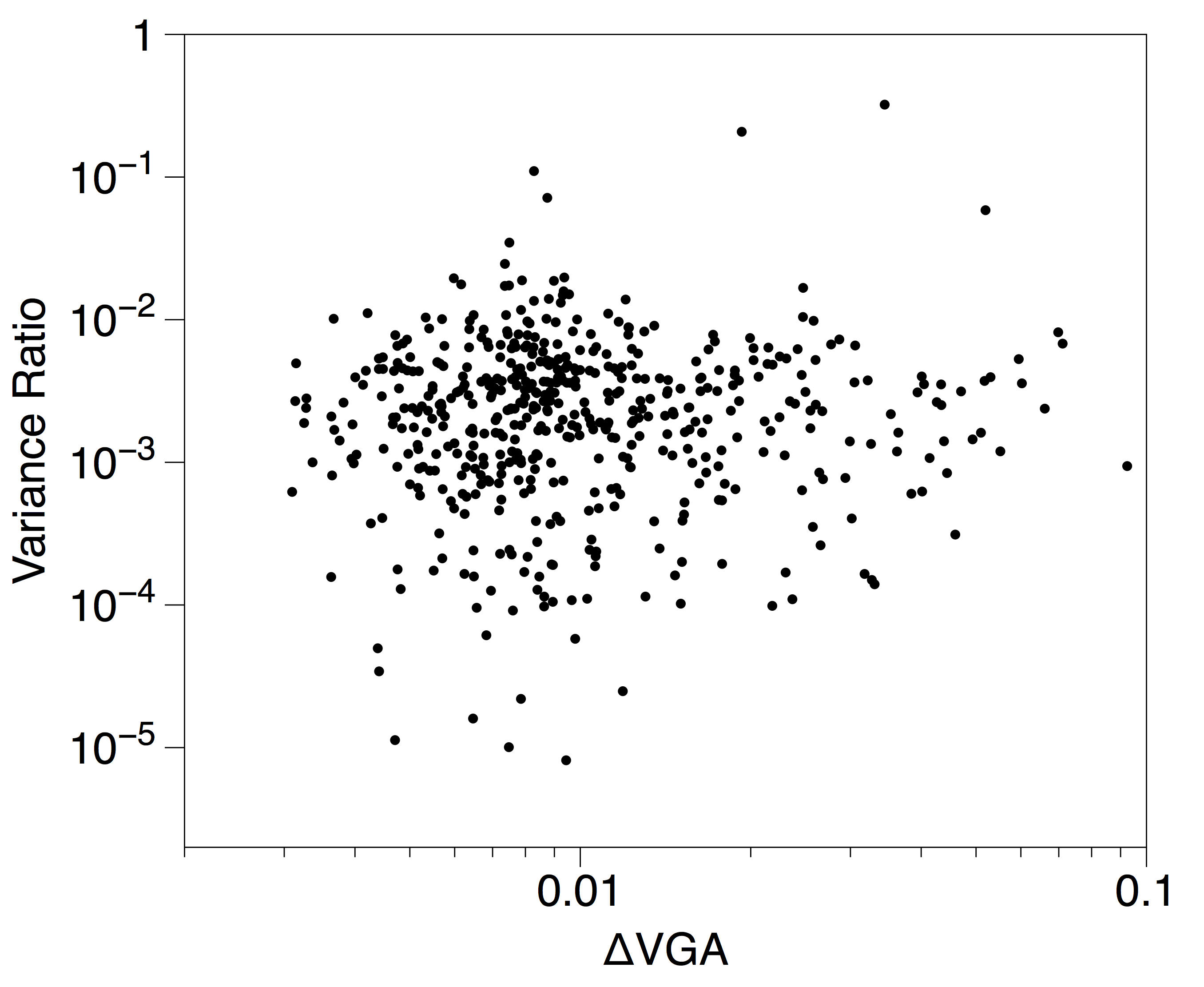}
\includegraphics[width=0.43\columnwidth]{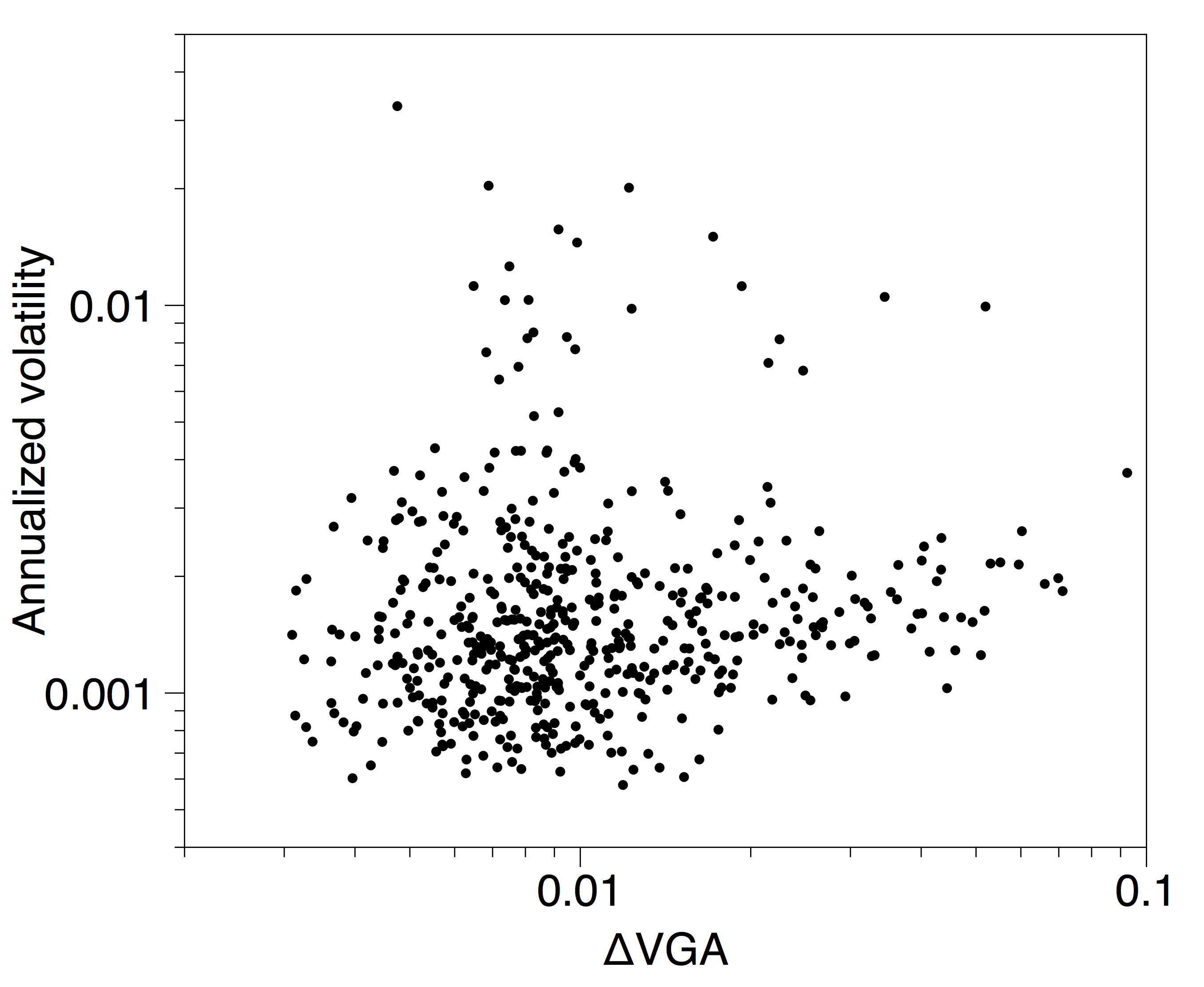}
\includegraphics[width=0.49\columnwidth]{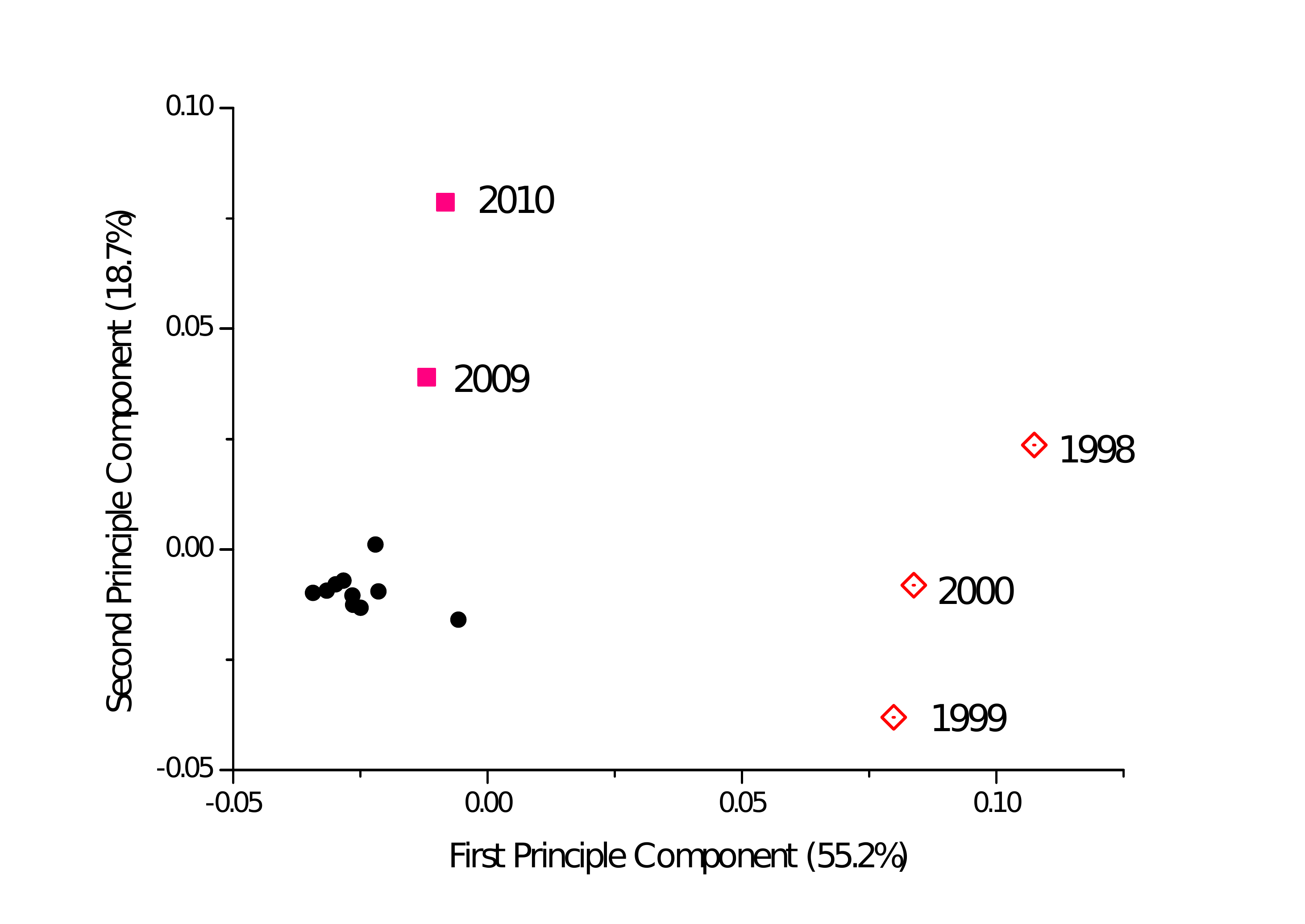}
\includegraphics[width=0.49\columnwidth]{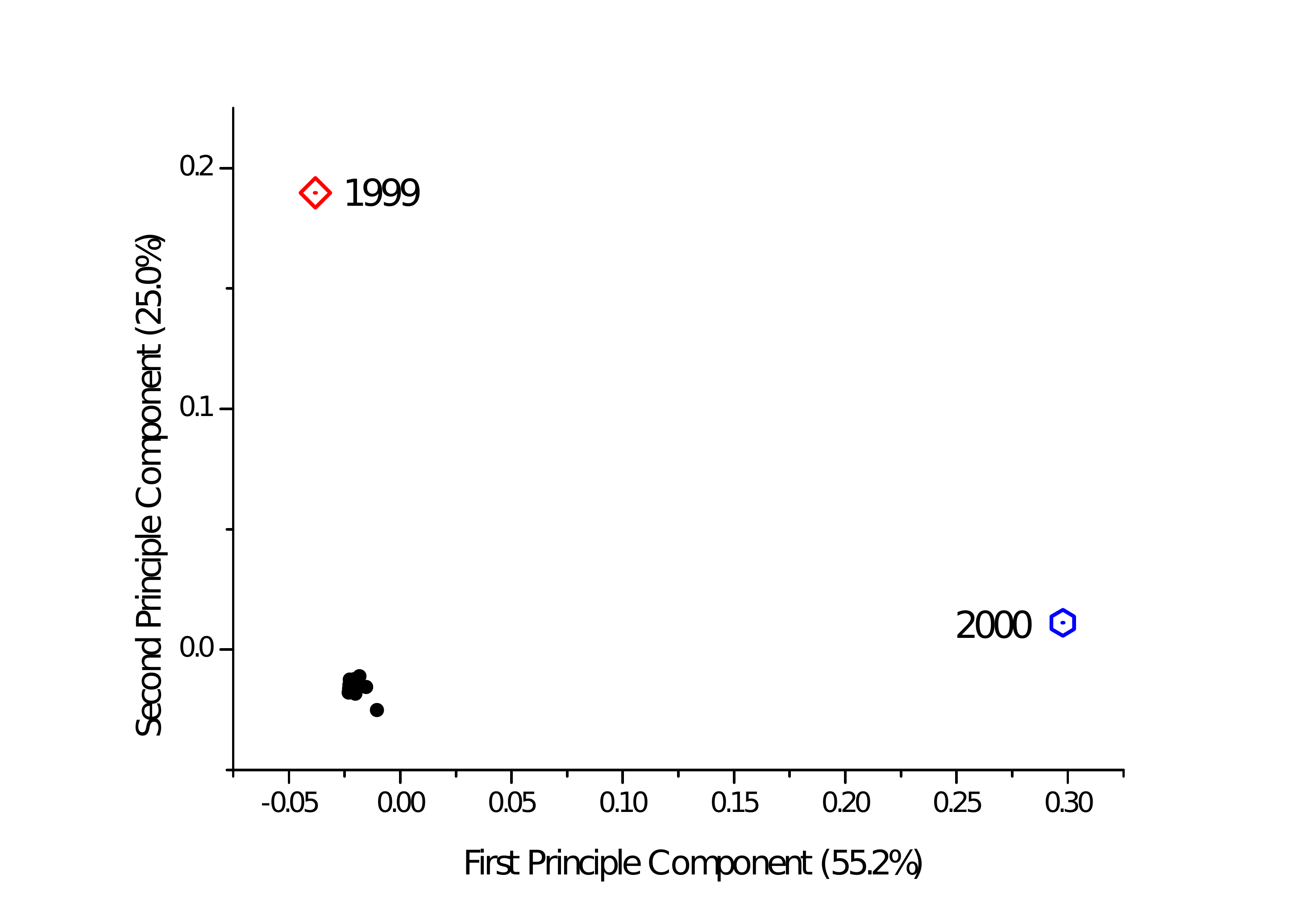}
\caption{(Upper panels) Scatter plot of the $\Delta$VGA (HVG) versus Variance Ratio (left panel) and annualized volatility (right panel) for each year in the period 1998-2012 for the set of 35 companies in the dataset ($N=525$ samples). No obvious correlation shows up (Pearson correlation $r=0.085$ and $0.032$ respectively). (Bottom, Left panel) PCA on the space $\{{d}^{\bf c}(\text{year})\}_{\text{year}=1998}^{2012}$ where $d$ is the distance between the top and bottom degree distributions (each point is a 35-dimensional vector describing the financial state of a given year in the PCA space spanned by the distance of top vs bottom degree distributions, for each of the 35 stock prices). Here we project points in the 2-dimensional space spanned by the first two principal components. We can see how three clusters emerge naturally: one agglomerating the years 1998,1999,2000 (which coincide with the .com bubble), one agglomerating 2009,2010 (coinciding with the period of the global financial crisis), and another cluster with the rest of the years (which coincide with periods of relatively stable financial activity).
%(Middle panel) Equivalent PCA projection on the space  $\{I_{\text{VG}}^{\bf c}(\text{year})\}_{\text{year}=1998}^{2012}$ where ${I_{\text{VG}}}^{\bf c}(\text{year})$ is now the visibility graph \emph{irreversibility} of a specified year, which is distance between the in and out degree distributions (Flanagan and Lacasa, 2016). Results are qualitatively similar. 
(Bottom, Right panel) Equivalent PCA projection on the space  $\{{|\log(VR)|^{\bf c}_{(\text{year})}}\}_{\text{year}=1998}^{2012}$, where the feature extracted from each stock price is now $|\log(VR)|$. We find that this feature is less informative. Accordingly, we therefore interpret that the difference between local maxima and local minima are not simply related to differences between marginal distributions, but a difference in the correlation (temporal ordering) structure.}
\label{PCAFinancial}
\end{figure*}

% That is to say, each year acts like a 'subject' . 
First, we show in the upper panels of Fig.\ref{PCAFinancial} that $\Delta$VGA is not correlated with the measure based on Variance Ratio (VR, left panel, Pearson correlation $r=0.085$), nor with the standard econometric measure to capture financial series fluctuations (annualized volatility, right panel, Pearson correlation $r=0.032$). This means that $\Delta$VGA is a genuinely different measure in this domain. The correlation between an index and its volatility is indeed related to the skewness of the index, hence this result further validates the hypothesis that $\Delta$VGA is independent of skewness.\\

Then, we perform principal component analysis (PCA) on  the vector $\{{d}^{\bf c}(\text{year})\}_{\text{year}=1998}^{2012}$  where ${d}^{\bf c}(\text{year})$ is the $\Delta$VGA (HVG) for company $\bf c$ for a specified year, and project results on a two-dimensional space spanned by the first two principal vectors of the PCA projection. This plot is shown in the left panel of Figure \ref{PCAFinancial}, 
%with a comparison in the middle panel of the same figure to a previous calculation which computes the \emph{visibility graph irreversibility} instead (which is the distance between the in and out degree distributions of the top HVG, see Flanagan and Lacasa, {\it Physics Letters A} 2016). The results are qualitatively similar, namely 
and indicates that the measure based on $\Delta$VGA is informative and we can cluster periods of financial turmoil together (the global financial crisis 2009-2010 is found separated from the dot-com bubble 1998-2000 and from the rest of the years).
In the right panel of the same figure, we further compare equivalent plots produced via $|\log(VR)|$. We thus compute $|\log(VR)|$ for each year, for each company, and perform PCA on the space $\{{|\log(VR)|^{\bf c}_{(\text{year})}}\}_{\text{year}=1998}^{2012}$. In this case no robust clustering appears.\\
%In Figure \ref{ExampleSeries} we plot $\{{d}^{\bf c}(\text{year})\}_{\text{year}=1998}^{2012}$ for two companies; AIG (American International Group) and C (Citigroup). We observe that AIG (American International Group) shows a large distance between the top and bottom degree distributions for the years 1998, 1999 and 2000 (corresponding to the dot-com bubble), then decreasing for the subsequent years. This pattern was also observed in many of the other companies. On the other hand, C (Citigroup) also shows a large spike in the years 2009 and 2010 (corresponding to the peak of the financial crisis). The fact that companies are showing different evolutions of $d$ over the time period, together with the PCA results in Figure \ref{PCAFinancial}, could hint to us that $d$ is a meaningful statistic in terms of financial series.

%\begin{figure*}
%\includegraphics[width=0.5\columnwidth]{AIG.pdf}
%\includegraphics[width=0.49\columnwidth]{C.pdf}
%\caption{Plot of $\{{d}^{\bf c}(\text{year})\}_{\text{year}=1998}^{2012}$ for two companies; AIG (American International Group) in the left panel, and C (Citigroup) in the right panel. The two companies show $d$ evolving differently, with Citigroup exhibiting a spike during the peak of the financial crisis (2009 to 2010).}
%\label{ExampleSeries}
%\end{figure*}

The interpretation of these results is as follows: the local fluctuations in the time arrangement of local maxima differs from the arrangement of local minima, and these differences vary with respect to the overall state of financial stress, i.e. collectively the set of stock prices is responsive to the financial stress state, and the difference between peak and pit statistics is informative of such collective financial state. As a result, it is easy to cluster apart years where the financial system was 'in equilibrium' from periods where the financial system is under stress and driven out of equilibrium, such as periods of financial bubbles or global crisis, which emerge as different periods in PCA space. We find that the statistic $|\log(VR)|$ fails to capture such traits: this is, according to our previous theoretical analysis, indicative that the difference between local maxima and local minima which cluster apart financial periods are not simply related to differences between marginal distributions, but rely on more subtle differences in the correlation (temporal ordering) structure.

\subsection{\textsf{Application: global daily temperature time series}}
\begin{figure}
\includegraphics[width=12 cm]{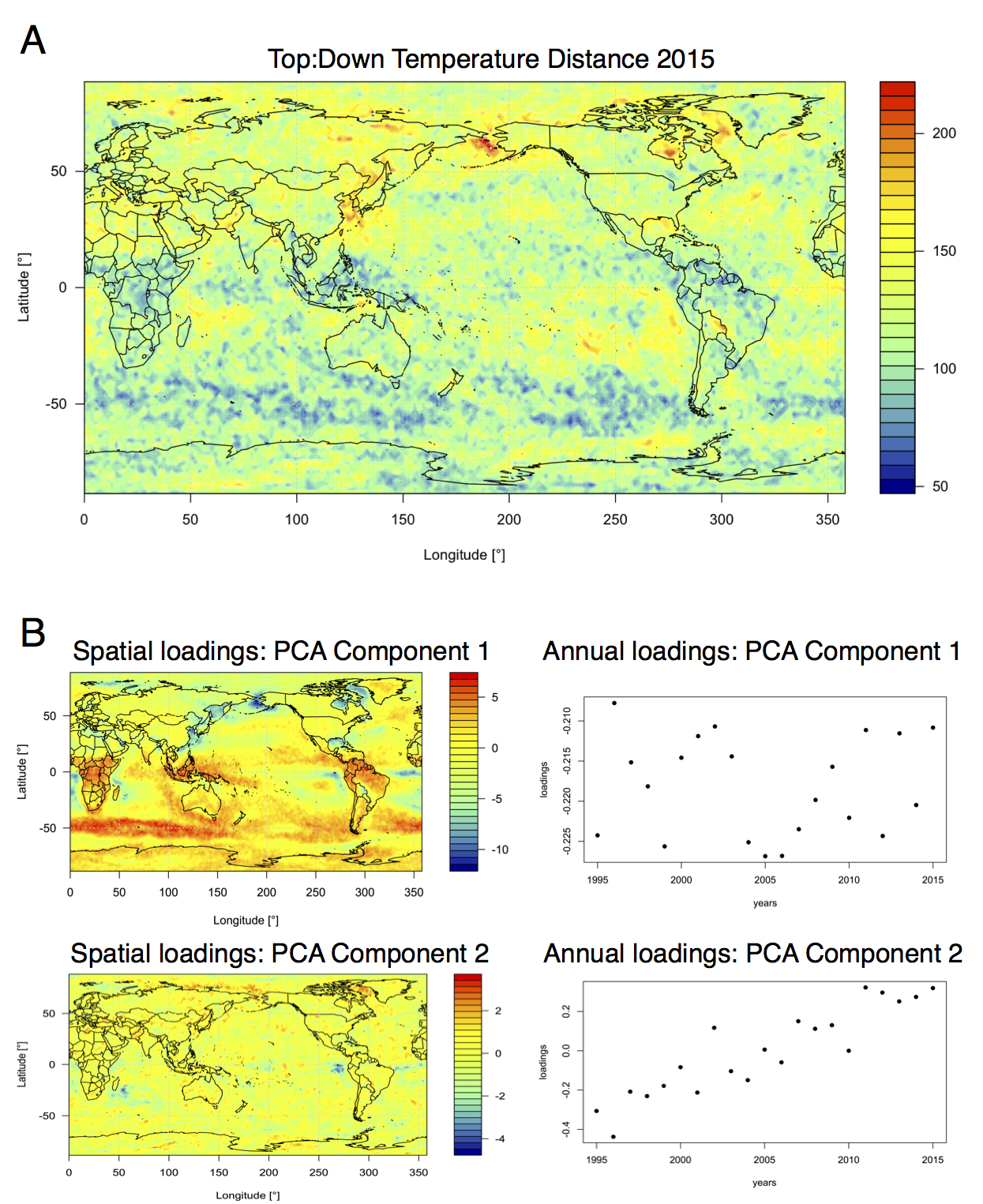}
\caption{Panel A: Distribution of Top:Down Distance values calculated from yearly temperature time series in 2015, with warmer colors indicating greater difference between top and bottom visibility graphs. Panel B: results of PCA applies to Top:Down Distance maps for the years 1995-2015. The first component accounted for around 40\% of the variance with stable yearly loadings. Component two accounted for around 5\% of the variance and showed marked changes in yearly loadings over time.}
\label{fig:TopDownMeteo}
\end{figure}

The $\Delta$VGA map for the year 2015 is seen in Panel A of \ref{fig:TopDownMeteo}. As a first step we quantified the relation between $\Delta$VGA and several moments of the temperature distribution. To this end, we derived $\Delta$VGA spatial heat maps for each of the years 1995-2015, and we then calculated the correlation between the observed value of $\Delta$VGA (per grid point) and the following parameters of the yearly temperature time series: (i) mean, (ii) standard deviation, (iii) skewness , and (iv) kurtosis. Note that this returns a set of pair-wise correlation values \emph{per year} and we can then examine the distribution of these correlation values across years. Similarly to what we found for the validation case, the correlation with mean was low (Mean Pearson's R; $r = .06 \pm .03$; the correlation with standard deviation, r = 0.24 $\pm .05$; with skewness, r = -0.09, $\pm 0.02$ and with kurtosis, r = -0.13, $\pm 0.04$.  Note the low correlation with skewness, which strongly suggests that $\Delta$VGA measure is not loading on extreme events \cite{Amor}. The correlation between $\Delta$VGA and the standard deviation was moderate, and the correlations with mean and kurtosis minimal.  This suggests that $\Delta$VGA provides information not captured by typical moments.\\
A Principal Component Analysis applied to the 21 Distance maps (1995-2015) identified a first component that accounted for 40\% of the variance, and a second that accounted for 5\% (see \ref{fig:TopDownMeteo}). The yearly loadings on the first component were quite stable across the years. In contrast, the yearly loadings on the second component show a strong linear change with time. 

\section{\textsf{Discussion}}
In this work, we first presented the theoretical and practical motivations for developing new methods for studying detailed features of local minima and maxima in various temporal domains. We provide analytical arguments that suggest that such issue can be addressed by a combinatorial method {--originally studied in the context of solar activity \cite{previous}--} based on deriving two visibility graphs from a single time series. We have thoroughly validated the concept by proposing a detailed methodology which we apply on a battery of synthetic time series extracted from complex dynamics of different origin (encompassing correlated stochastic processes, chaotic dynamics and processes involving multiple scales). We have shown, both numerically and rigorously, that exploiting the difference and asymmetry between top and bottom VG/HVG (what we labeled as $\Delta$VGA) correctly identifies the peak/pit asymmetries and outperforms standard methods in several cases.\\
In applying this method, we first found that it offered a new view into spontaneous resting state dynamics in the human brain. While prior neuroimaging work based on amplitude-variance-asymmetry had pointed to sensory cortices as ones having different peak vs. pit dynamics \cite{AVA}, the current results identify frontal regions as exhibiting asymmetric resting-state BOLD fluctuations during wakeful rest. Furthermore, these asymmetric patterns were also found during the deeper sleep stages (N2 and N3), which is a departure from prior findings (DAVIS16) where AVA failed to identify such signatures. An important result was that in all clusters, these dynamics were driven by differences in the relative frequency of time points (nodes) with relatively low degree, indicating that these differences are \emph{not} due to a difference in relative frequency of rare, extreme events but due to differences in frequency of time points with relatively moderate connectivity -- i.e., a very local phenomenon.\\ 
Then, we were able to cluster periods of financial activity according to differences in peak/pit asymmetry via principal component analysis in visibility graph feature space. Clusters gathering periods of similar financial stress emerge naturally, and capture finer-grained structure than a basic analysis based on AVA. Finally, we demonstrated a straightforward application of the methods in question to daily temperature time series. The latter produced two main results. First, already hinted at by our analysis of synthetic series, there were either very modest or no correlations between $\Delta$VGA and typically studied distribution moments including mean, standard deviation, skewness or kurtosis. Second, A PCA analysis revealed well structured spatio-temporal correlations between changes in $\Delta$VGA in several latitude bands, broadly matching planetary-scale topology of the sub-tropical and polar jet streams. The yearly PCA loadings of the main component did not identify a strong cyclical pattern consistent with El-Nino events, but the loadings on component 2 suggest a gradual monotonic change in $\Delta$VGA correlations over time between several areas.\\

\noindent In addition, we have shown that $\Delta$VGA goes beyond basic statistics defined over VG/HVG since topological features of a standard VG/HVG cannot capture accurately the situation when the signal is aperiodic and alternates between large and small values (i.e. the structure of small data is completely lost as they will always be nodes with degree $k=2$), such being for instance the case of a chaotic process with a disconnected two-band chaotic attractor among other cases.\\

\noindent We anticipate that $\Delta$VGA will provide a efficient feature extraction method which will be valuable for statistical learning analysis across the disciplines. Finally, note that in this work the definitions of $\Delta$VGA mainly reduce to a comparison between global topological features. However the same methodology also allows to extract local topological features (e.g. motifs \cite{motifs}), whose comparison can thus be exploited for early warning and anomaly detection purposes.\\

% \textcolor{red}{LL: these are old comments:\\
% - Other features from top/bottom VG exploiting convexity vs ordering\\
% - Sleep with regard to fMRI Convexity and fMRI % Ordering\\ 
% - CDF vs Count}

%\bibliography{bib_vis}

%This defines the bibliographies style. Search online for a list of available styles.
%\bibliographystyle{abbrv}

\section{\textsf{APPENDIX: Proofs of the theorems}}

\noindent {\bf Theorem 1.} {\it Let $\{x(t)\}$ be a bi-infinite time series generated by (i) white noise with a continuous probability density $f(x)$ or by (ii) symmetric correlated red noise. Then $\Delta$\text{VGA}$=0$ for HVG.}\\

\noindent {\bf Proof. }The proof for this theorem is straightforward. Firstly, we consider the case (i). We rely on a result of \cite{Lacasa2009} where it has been proved that the degree distribution of the HVG of $\{x(t)\}$ generated by white noise is
\begin{equation}
P(k)=(1/3)(2/3)^{k-2}
\label{dd}
\end{equation}
Trivially, if $\{x(t)\}$ is white noise extracted from $f(x)$, this means that it is a collection of i.i.d. random variables extracted from $f(x)$. But then $\{-x(t)\}$ is again a collection of i.i.d. variables extracted from a different density $g(x)$. By virtue of a theorem proved in \cite{pre}, the degree distribution of the HVG associated to such series coincides with eq.\ref{dd}, hence $\Delta$VGA$=0$. This closes the first part of the proof.\\

\noindent For (ii), we only need to remark that red noise (AR(1)) is generated via the stochastic map
\begin{equation}
x_t=r\cdot x_{t-1}+\xi, \quad \xi \in N[0,1] \nonumber
\label{AR1}
\end{equation}
Trivially, if $\xi$ is drawn from a symmetric distribution (such as the Gaussian function above), then both
$\{x(t)\}$ and $\{-x(t)\}$ are equally likely realizations of eq. \ref{AR1}, hence $\Delta$VGA vanishes. \qed \\
\begin{figure}
\centering
\includegraphics[width= 12 cm]{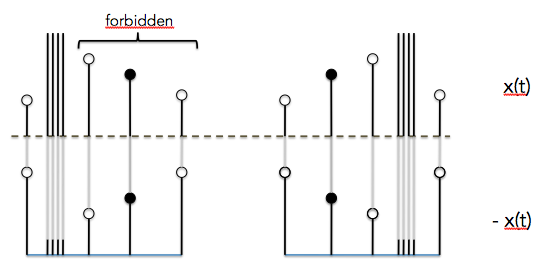}
\caption{Schematic representation of the two sets of diagrams contributing to $P^{\text{bot}}(3)$ (which is equivalent to $P^{\text{top}}(3)$ as computed on $-x(t)$). In every case the reference node $x_0$ is highlighted as black solid dot. Hidden nodes (an arbitrary large amount of them) are schematized as vertical bars with no dots on top. The first diagram (bottom, left) does not actually appear in $\{-x(t)\}$ as the associated diagram in the original series is forbidden (in the fully chaotic logistic map we never find three consecutive data points in decreasing order \cite{nonlinearity}). Accordingly, the only set of diagrams is the one pictorically represented in the bottom right. The relative ordering of the data in the original chaotic time series is represented in the upper part.}
\label{fig:proof2}
\end{figure}

\noindent {\bf Theorem 2.} {\it Let $\{x(t)\}$ be a bi-infinite time series generated by a fully chaotic logistic map. Then $\Delta$\text{VGA}$>0$ for HVG.}\\

\noindent {\bf Proof. }The proof for this theorem is somewhat more convoluted than for theorem 1. Since $\Delta$VGA is semi-positive definite, it vanishes if and only if $\forall k=2,3,\dots: P^{\text{top}}(k)=P^{\text{bot}}(k)$. Thus, in order to prove positivity we will only need to find that for a given degree $k=m$, $P^{\text{top}}(m)\neq P^{\text{bot}}(m)$. After a quick numerical inspection, we choose $m=3$. In \cite{toral} it was analytically proved that for a fully chaotic logistic map, $P^{\text{top}}(3)=1/3$. Without loss of generality, we will prove that $P^{\text{bot}}(3)<1/3$.\\

\begin{figure}
\centering
\includegraphics[width= 8 cm]{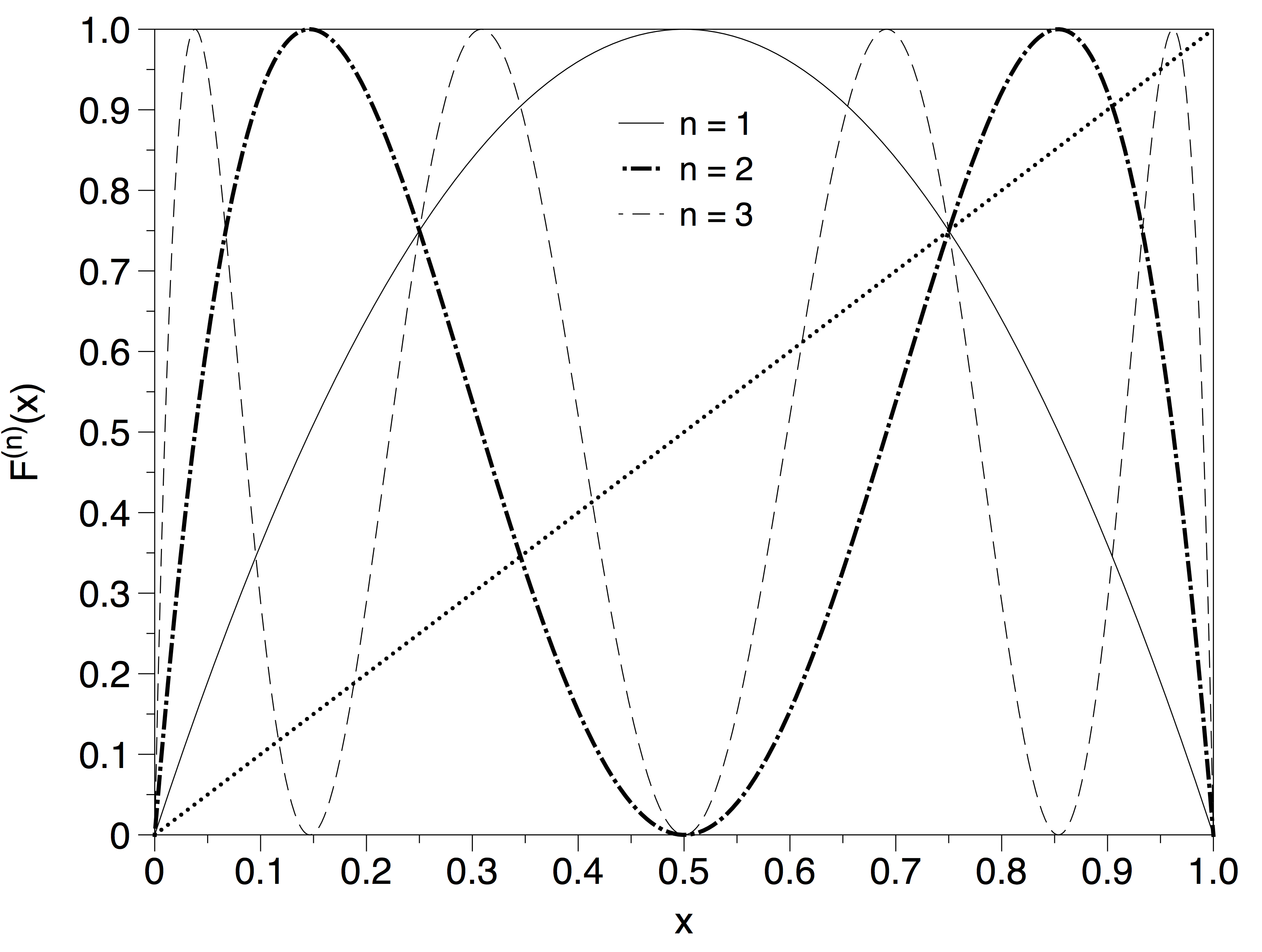}
\caption{A few iterations of the fully chaotic logistic map $x(t+1)=F(x(t));\ F(x)=4x(1-x)$}
\label{fig:cobweb}
\end{figure}

\noindent We capitalise on the perturbative expansion formalism advanced in \cite{nonlinearity} to analytically compute the degree probabilities of HVG associated to chaotic maps with smooth physical invariant measure. This expansion is diagrammatic, and in particular it expands on the number of hidden nodes (see \cite{nonlinearity} for details). In order to apply this technique here, we first observe that the bottom HVG of $\{x(t)\}$ is equivalent to the top HVG of $\{-x(t)\}$. Accordingly, the set of diagrams that comply with $P^{\text{bot}}(3)$ in the original series is schematized in figure \ref{fig:proof2}. It is easy to see that this is an infinitely unfolding combination of diagrams (each with a different number of hidden nodes), and as such formally we have
$$P^{\text{bot}}(3)= \sum_{\alpha=0}^{\infty} P^{\text{bot}}_{\alpha}(3),$$
where $\alpha$ is the number of hidden nodes in each diagram. Let us denote $F(x)=4x(1-x)$ as the fully chaotic logistic map, $f(x)=\pi^{-1}x^{-1/2}(1-x)^{-1/2}$ as the invariant measure of the map. Since this is a Markovian and deterministic map, the probability of observing a datum $x$ given that the previously observed datum is $y$ is simply $\delta(x-F(y))$, where $\delta(\cdot)$ is the Dirac-delta distribution. The integral corresponding to each contribution to the probability is then:
\begin{eqnarray}
&&P^{\text{bot}}_{\alpha}(3)=\int_0^{x_0} f(x_{-1})\delta(x_0 - F(x_{-1})) dx_{-1}\int_0^1 dx_0 \int_{x_0}^1 \delta(x_1-F(x_0))dx_1 \cdot \\ \nonumber \cdot 
&& \bigg[ \prod_{i=1}^\alpha \int_{x_1}^1 \delta(z_i - F^{(i+1)}(x_0))dz_i\bigg] \int_0^{x_0}\delta(x_2-F^{(\alpha +2)}(x_0)) dx_2  
\end{eqnarray}
The first important point to note is that the first integral can be computed using the scaling properties of the Dirac delta:\\
$$\int_0^{x_0} f(x_{-1})\delta(x_0 - F(x_{-1})) dx_{-1} = \sum \frac{f(x^*)}{|F'(x^*)|},$$
where $x^*$ are the roots of the equation $F(x)=x_0$ (in particular, we only need to sum up over those roots that belong to the interval of integration, that is, only those roots that satisfy $x^*<x_0$. For $x_0 \in [0,3/4]$, the only root which fulfils such inequality is $x^*=(1-\sqrt{1-x_0})/2$, hence
\begin{equation}
\int_0^{x_0} f(x_{-1})\delta(x_0 - F(x_{-1})) dx_{-1} = \frac{f(x^*)}{|F'(x^*)|}=\frac{1}{\pi\sqrt{(1-\frac{\sqrt{1-x_0}}{2})(1-\frac{1-\sqrt{1-x_0}}{2})}4\sqrt{1-x_0}}=:W(x_0).
\label{Wx}
\end{equation}
The rest of the integrals only have the effect of reducing the interval of integration of $x_0$, as these are Dirac integrals and thus are either one or zero, since $\int_a^b \delta (x-y)dx$ is one if $y\in[a,b]$ and zero otherwise.\\
Let us assume for a moment that we remove in the integral above any contribution of nodes above $x_1$. After a little algebra, this particular case labeled $Q$ would yield a probability
\begin{eqnarray}
&&Q=\int_0^{x_0} f(x_{-1})\delta(x_0 - F(x_{-1})) dx_{-1}\int_0^1 dx_0 \int_{x_0}^1 \delta(x_1-F(x_0))dx_1 =\nonumber  \\ 
&&=\int_0^{3/4} \frac{f(x^*)dx_0}{4\sqrt{1-x_0}}=\int_0^{3/4} W(x_0)dx_0=1/3=P^{\text{top}}(3),
\label{xxx}
\end{eqnarray}
where $W(x_0)$ was defined in eq.\ref{Wx}, and the shrinking of the interval of integration in $x_0$ from $[0,1]$ to $[0,3/4]$ comes from the integral $\int_{x_0}^1 \delta(x_1-F(x_0))dx_1$, which is only equal to one if $F(x_0)>x_0$, what happens for $x_0 \in [0,3/4]$ (see figure \ref{fig:cobweb}). Importantly, any other additional Dirac delta integral (associated to the contribution of further nodes --hidden or otherwise--) of the form $\int_a^b \delta (x-y)dx$ will have the effect of {\it further shrinking} the interval $[0,3/4]$ in the integral over $x_0$, in such a way that for each order $\alpha$ the original interval of integration of $x_0$ shrinks into a subinterval ${\cal I}_\alpha$. To showcase such situation let us consider in full detail the computation of $P^{\text{bot}}_0(3)$ and $P^{\text{bot}}_1(3)$.\\

\noindent \underline{$\alpha=0.$}\\
 $$P^{\text{bot}}_0(3)=\int_0^{x_0} f(x_{-1})\delta(x_0 - F(x_{-1})) dx_{-1}\int_0^1 dx_0 \int_{x_0}^1 \delta(x_1-F(x_0))dx_1\int_0^{x_0}\delta(x_2-F^{(2)}(x_0)) dx_2  $$
Let's consider the last two integral terms above: they require\\
$F(x_0)>x_0$ and $F^{(2)}(x_0)<x_0$. Looking at figure \ref{fig:cobweb} and considering the fixed point structure of $F(x)$ and higher iterates of the map, we find that the first condition shrinks the interval of integration of $x_0$ from $[0,1]$ to $[0,3/4]$, whereas the second condition is fulfilled for $x_0 \in [\frac{5-\sqrt{5}}{8},\frac{3}{4}]$. The intersection of the two conditions above yields an interval $${\cal I}_0=[\frac{5-\sqrt{5}}{8},\frac{3}{4}]$$, and therefore
$$P^{\text{bot}}_0(3)=\int_{\frac{5-\sqrt{5}}{8}}^{3/4}W(x_0)dx_0\approx 0.1333$$

\begin{figure}
\centering
\includegraphics[width= 9 cm]{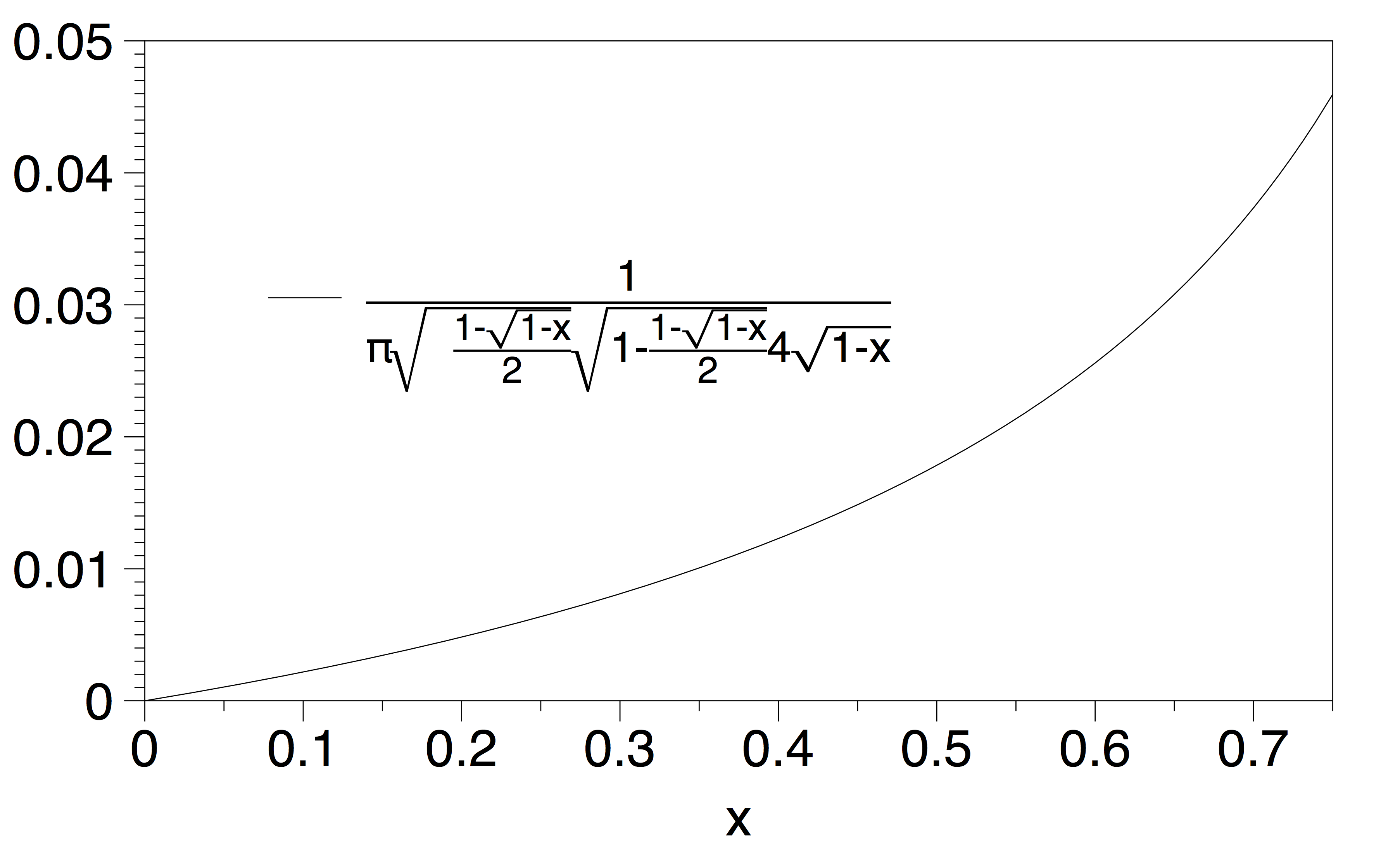}
\caption{Integrand in the last integral of Eq.\ref{xxx}.}
\label{proof1}
\end{figure}

\noindent \underline{$\alpha=1.$}\\
 $$P^{\text{bot}}_1(3)=\int_0^{x_0} f(x_{-1})\delta(x_0 - F(x_{-1})) dx_{-1}\int_0^1 dx_0 \int_{x_0}^1 \delta(x_1-F(x_0))dx_1\int_{x_1}^1\delta(z_1-F^{(2)}(x_0))dz_1 \int_0^{x_0}\delta(x_2-F^{(3)}(x_0)) dx_2  $$
Let's then consider the last {\it three} integral terms above: they require\\
$F(x_0)>x_0$, $F^{(2)}(x_0)> F(x_0)$, $F^{(3)}(x_0)<x_0$. Looking at figure \ref{fig:cobweb} and considering again the fixed point structure of $F(x)$ and higher iterates of the map, we find that the first condition shrinks the interval of integration of $x_0$ from $[0,1]$ to $[0,3/4]$, whereas the second condition is fulfilled now for 
$x_0 \in [0,1/4]\cup [3/4,1]$. The third condition is fulfilled in 
$x_0 \in [0.11698,0.188255]\cup [0.41318, 0.61126]\cup[3/4, 0.95048]\cup[0.96985,1]$. The intersection of all these conditions thus provides $${\cal I}_1=[0.11698,0.188255]$$, and therefore
$$P^{\text{bot}}_1(3)=\int_{0.11698}^{0.188255}W(x_0)dx_0\approx 0.03174$$
One can proceed indefinitely adding higher orders in $\alpha$, which yield smaller and smaller contributions to the total probability. 
In other words, for any $\alpha >0$ the resulting interval ${\cal I}_\alpha$ is always a subset of $[0,3/4]$. Since the integrand in the last integral of eq.\ref{xxx} is always positive (see figure \ref{proof1} for a graphical support), we conclude that $P^{\text{bot}}_\alpha (3)<Q$. Furthermore, we need to prove that if we concatenate the subinterval obtained for each $\alpha$, its union will always be smaller than $[0,3/4]$: 
\begin{equation}
\bigcup_{\alpha=0}^\infty {\cal I}_{\alpha} \subset [0,3/4]
\label{key}
\end{equation}
This last statement can be easily proved by showing that there are always subintervals of $[0,3/4]$ not covered in any ${\cal I}_\alpha$. In particular, consider the subinterval $[1/4,(5-\sqrt{5})/8]$. We have explicitely shown that such subinterval is not included in ${\cal I}_0$, nor in ${\cal I}_1$. Now, for $\alpha \geq 1$ by construction one of the conditions which is always present is $F^(2)(x_0)>F(x_0)$. Now, such condition is fulfilled for $x_0 \in [0,1/4]\cup[3/4,1]$, leaving our interval $[1/4,(5-\sqrt{5})/8]$ out. As this situation holds $\forall \alpha$, this directly yields eq.\ref{key}.Finally, since integration is monotonic and since the integrand $W(x_0)$ is positive, we trivially have
$${\cal A}\subset {\cal B}\Rightarrow \int_{\cal A}W(x_0)dx_0 < \int_{\cal B}W(x_0)dx_0$$

Altogether, this means that $P^{\text{bot}}(3)= \sum_{\alpha=0}^{\infty} P^{\text{bot}}_{\alpha}(3)<Q=P^{\text{top}}(3)$, what concludes the proof. \qed\\

\noindent \textbf{Acknowledgments. } We wish to thank Dr. Alon Angert for his advice on meteorological time series.  UH`s work was conducted in part while serving at and with support of the National Science Foundation. Any opinions, findings, and conclusions or recommendations expressed in this material are those of the author(s) and do not necessarily reflect the views of the NSF. LL's acknowledges funding from an EPSRC Early Career Fellowship EP/P01660X/1.\\

\noindent \textbf{Author contributions. } UH and LL designed research; UH, BD, JI, RF and LL conducted research, ET and HL contributed data, all authors discussed results and UH, JI, RF and LL wrote the paper.\\

\noindent \textbf{Competing financial interests. }None.


\end{document}